%% file: main.tex
\documentclass[11pt,a4paper]{article}

\usepackage[margin=28mm]{geometry}
\usepackage{amsmath,amssymb,amsthm,mathtools,mathrsfs}
\usepackage{thmtools}
\usepackage{array,booktabs,longtable,tabularx}
\usepackage{enumitem}
\usepackage[numbers,sort&compress]{natbib}
\usepackage[pagebackref=true]{hyperref}
\usepackage{cleveref}
\usepackage{microtype}
\usepackage{lineno}
\usepackage{authblk}

\renewcommand*{\backref}[1]{}
\renewcommand*{\backrefalt}[4]{%
  \ifcase #1 %
  \or (Cited on page~#2.)%
  \else (Cited on pages~#2.)%
  \fi
}

\allowdisplaybreaks
\setlist{itemsep=2pt,topsep=4pt}

\newcommand{\Pow}{\operatorname{Pow}}
\newcommand{\Comp}{\operatorname{Comp}}
\newcommand{\boxi}{\operatorname{box}}

\newcommand{\C}{\mathcal C}
\newcommand{\sd}{\operatorname{sd}}
\newcommand{\depth}{\operatorname{depth}}
\newcommand{\Nerve}{\operatorname{Nerve}}
\newcommand{\Z}{\mathbb Z}
\newcommand{\F}{\mathbb F}

\newcommand{\G}{\Gamma}

\newcommand{\ang}[1]{\langle #1 \rangle}
\newcommand{\kk}{\mathsf{K}}
\newcommand{\dd}{D_3^{*r}}
\newcommand{\Pt}{\operatorname{P}}

\newcolumntype{Y}{>{\raggedright\arraybackslash}X}

\usepackage[colorinlistoftodos]{todonotes}

\theoremstyle{plain}
\newtheorem{theorem}{Theorem}[section]
\newtheorem{lemma}[theorem]{Lemma}

\newtheorem{observation}[theorem]{Observation}
\newenvironment{claimproof}{\par\noindent\underline{Proof of Claim:}}{\hfill $\blacksquare$}
\newtheorem{claim}[theorem]{Claim}
\newtheorem{proposition}[theorem]{Proposition}
\newtheorem{corollary}[theorem]{Corollary}
\theoremstyle{definition}
\newtheorem{definition}[theorem]{Definition}

\theoremstyle{remark}
\newtheorem{remark}[theorem]{Remark}

\author[1]{L. Sunil Chandran%
  \thanks{Email: \texttt{sunil@iisc.ac.in}}}

\author[1]{Jinia Ghosh%
  \thanks{Email: \texttt{jiniaghosh@iisc.ac.in}}}

\author[1]{Suraj Kumar Sahoo%
  \thanks{Email: \texttt{surajks@iisc.ac.in}}}

\affil[1]{Indian Institute of Science, Bangalore, India}

\date{}

\title{Boxicity of Power Graphs of Finite Nilpotent Groups}

\begin{document}

\maketitle

\begin{abstract}
The \textit{boxicity} of an undirected graph $\Gamma$, denoted as $\operatorname{box}(\Gamma)$, is the smallest non-negative integer $d$ such that $\Gamma$ can be represented as the intersection graph of axis-parallel rectangular boxes in $\mathbb{R}^d$. The \emph{power graph} of a group $G$, denoted as $\operatorname{Pow}(G)$, is an undirected graph whose vertex set is the group itself, with two elements being adjacent if one is a power of the other. 

We determine the exact boxicity of power graphs of finite nilpotent groups. For a cyclic group of order $n=\prod_{i=1}^{r}p_i^{e_i}>1$, where the $p_i$ are distinct primes, we prove that its power graph has boxicity $\sum_{i=1}^{r}e_i-\max_i e_i$. For a finite nilpotent group $G$, let the number of noncyclic odd Sylow subgroups be denoted by $t(G)$. If $G$ has odd order or if $G$ has a cyclic Sylow $2$-subgroup, then $\operatorname{box}(\operatorname{Pow}(G))=\operatorname{box}(\operatorname{Pow}(\mathbb{Z}_{\operatorname{exp}(G)}))+t(G)$, where $\operatorname{exp}(G)$ denotes the \emph{exponent} of a group $G$. When the Sylow $2$-subgroup is noncyclic, and the odd part is nontrivial, the boxicity equals this value or one more, and we characterise precisely when each of these values is attained. 
 
The boxicity of power graphs of arbitrary finite groups and the exact boxicity of the comparability graph of the Boolean lattice were posed as open questions in Discrete Mathematics, 349(7):115017, 2026. As a consequence of our results, we answer the latter question completely and the former question partially, by computing the exact boxicity of the power graphs of nilpotent groups.
\end{abstract}

\section{Introduction}
All graphs in this paper are finite, simple and undirected. A $d$-dimensional box representation of a graph is a function $f$ that maps the vertices of the graph to axis-parallel boxes in $\mathbb{R}^d$, such that two distinct vertices $u$ and $v$ in the graph are adjacent if and only if $f(u)\cap f(v)\neq \emptyset.$ 
\begin{definition}
     The boxicity of a graph $\Gamma$, denoted as $\boxi(\Gamma)$, is the minimum $d\geq 0$ such that there is a $d$-dimensional box representation of $\Gamma$. 
\end{definition}
By convention, we assume that a complete graph has boxicity $0$. Note that interval graphs are precisely the graphs that have boxicity at most $1$.

The notion of boxicity was introduced by Roberts~\citep{Roberts}, who showed that any graph on $n$ vertices has a $\lfloor n/2\rfloor$-dimensional box representation. Boxicity has found applications in areas such as ecology~\cite{10.1007/BFb0070404} and operations research~\cite{opsut1981fleet}. 

The boxicity of several important graph classes has been studied. It is known that boxicity can be arbitrarily large even for restrictive classes like chordal bipartite graphs~\cite{chandran2011chordal}, split graphs~\cite{cozzens1983computing}, line graphs~\cite{chandran2011boxicity,10.1007/978-3-319-12340-0_7} and strongly chordal graphs~\cite{leaf}. On the other hand, it is known that planar graphs and strictly chordal graphs have boxicity at most $3$~\cite{thomassen1986interval,leaf} and that planar bipartite graphs and outerplanar graphs have boxicity at most $2$~\cite{hartman1991grid,Schein}. The boxicity of asteroidal triple-free graphs, circular arc graphs, Kneser graphs and the complement of line graphs has also been studied \cite{bhowmick2010boxicity,bhowmick2011boxicity,caoduro2023boxicity,caoduro2023boxicityy}. For more on the boxicity of graphs, the reader may refer to~\cite{chandran2026survey}.  

 Adiga, Bhowmick, and Chandran \cite{adiga2011boxicity} have examined the relationship between poset dimension, an order-theoretic concept, and boxicity, prompting investigations into boxicity within subclasses of comparability graphs and poset dimensions in the corresponding partial orders \cite{chalermsook2013graph, chaplick2018grid}. Recently, Chandran and Ghosh~\cite{chandran2026boxicity} studied the boxicity of an algebraically defined subclass of comparability graphs, called \textit{power graphs}. Outside of power graphs, the only other study of the boxicity of an algebraically defined graph class has been that of zero-divisor graphs.~\cite{kavaskar2025bounds,chandran2026zero,chandran2026compressed,caoduro2026boxicity}. 
 
 The \emph{power graph} of a finite group $G$, denoted as $\Pow(G)$, is a graph defined on the group $G$ such that two group elements are adjacent in $\Pow(G)$ if and only if one is the power of the other. Power graphs were introduced by Chakrabarty, Ghosh and Sen~\cite{chakrabarty2009undirected}. For more on power graphs, the readers may refer to~\cite{abawajy2013power,kumar2021recent}.  

Chandran and Ghosh~\cite{chandran2026boxicity} studied $\Pow(G)$ when $G$ is a cyclic group. They showed the following.
\begin{theorem}[\cite{chandran2026boxicity}]\label{prevthm}
    Let $1\leq m_1\leq m_2\leq \ldots \leq m_d$ be positive integers and $d\geq 2$. Let $n=p_1^{m_1}p_2^{m_2}\dots p_d^{m_d}$, where $p_1,p_2,\ldots,p_d$ are pairwise distinct primes. Then 
    \begin{enumerate}
        \item[(a)] $\boxi(\Pow(\Z_n))\leq   m_1+m_2 + \cdots+m_{d-1}$.
        \item[(b)] When $d$ is odd, $\boxi(\Pow(\Z_n))\geq \sum_{1\leq j\leq \frac{d-1}{2}}m_{2j-1}+m_{d-1}$.
        \item[(c)] When $d$ is even, $\boxi(\Pow(\Z_n))\geq \sum_{1\leq j\leq \frac{d-2}{2}}m_{2j}+m_{d-1}$. 
    \end{enumerate}
\end{theorem}

Moreover, in \cite{chandran2026boxicity}, an open question was posed: "What is the exact value of the boxicity of the power graph of the cyclic group?" In this paper, we address this question. Additionally, the inquiry regarding the boxicity of power graphs of arbitrary groups was raised in \cite{chandran2026boxicity}. We provide a partial answer by determining the exact boxicity of the power graph associated with a finite nilpotent group. 

\subsection{Our Results}

\subsubsection{Exact boxicity of the power graph of a cyclic group}\label{sec:cyclic}

Chandran and Ghosh \cite{chandran2026boxicity} posed the question of determining the exact boxicity of power graphs of finite cyclic groups. In this paper, we address this question by establishing a lower bound for the boxicity that matches the upper bound provided in Theorem~\ref{prevthm}. We prove the following:

 \begin{theorem}\label{thm: cyclic}
     Let $n=p_1^{m_1}p_2^{m_2}\dots p_d^{m_d}$, where $p_1,p_2,\ldots,p_d$ are pairwise distinct primes and $d \geq 2$. Then $\boxi(\Pow(\Z_n))= \sum_{i=1}^dm_i -\max_{i\in [d]} m_i$.
 \end{theorem}

 \subsubsection{Exact boxicity of the power graph of a nilpotent group}

Our second result is the exact boxicity of the power graph of any nilpotent group (\Cref{thm:main}). To state the result, we need to give a few notations and definitions. Let $G$ be a finite nilpotent group. It is well known that the Sylow subgroups of $G$ are normal and that $G$ can be written as a direct product of its Sylow subgroups, i.e, 
\begin{equation}\label{eq: nilpotent}
    G=\prod_{\substack{p \text{ is a prime factor of } |G|}}S_p, \text{ where $S_p$ is the Sylow $p$-subgroup of $G$}.
\end{equation}
Let $s(G)=\bigl|\{p:S_p\text{ is noncyclic}\}\bigr|$ and $t(G)=\bigl|\{p\text{ odd}:S_p\text{ is noncyclic}\}\bigr|$. 
Note that, when $|G|$ is odd, we have $t(G)=s(G)$.

For a finite group $H$, we define the function $ \exp(H)$ to be the exponent of $H$, i.e., the smallest positive integer $k$ such that $h^k=e$, for every $h \in H$. We write $\C(H)$ to denote the poset of cyclic subgroups of a finite group $H$, ordered by inclusion. If $H$ is a $p$-group, then the cover graph of $\C(H)$ is a rooted tree whose
root is the trivial subgroup. We prove this later in \Cref{lem:p-tree}. 
We need the following definition to state our results.
\begin{definition}\label{def:fork-depth}
Let $T$ be a finite \emph{rooted tree poset} (see Definition~\ref{def: rooted tree poset}). For $u\in T$, let $\depth(u)$ be the distance of $u$ from the root in the cover graph of $T$. For $k\ge2$, define $h_k(T)=\max\bigl\{\depth(u):u\text{ has }k\text{ pairwise incomparable strict descendants in $T$}\bigr\}.$ The maximum of the empty set is $-\infty$.  For a $p$-group $H$, we write
$h_k(H)=h_k(\C(H))$.
\end{definition}

The descendants in this definition need not cover $u$. The following theorem is our second main result.

\begin{theorem}\label{thm:main}
Let $G$ be a finite nilpotent group as expressed in \Cref{eq: nilpotent}. 
Let $n=\exp(G)$. 
\begin{enumerate}[label=\textup{(\roman*)}]
\item If $S_2$ is absent or cyclic, then $\boxi(\Pow(G))=\boxi(Pow(\Z_n))+t(G).$

\item Suppose $S_2$ is noncyclic and the odd part of $G$ is nontrivial.  Let \(\exp(S_p)=p^{e_p},\ a=e_2,\ m=\max_{p \neq 2}e_p,\ q=\min\{a,m\}.\) 
Then
$$
 \boxi(\Pow(G))=
 \begin{cases}
  \boxi(\Pow(\Z_n))+t(G)+1,&h_3(S_2)\ge q-1,\\[1mm]
  \boxi(\Pow(\Z_n))+t(G),&h_3(S_2)\le q-2.
 \end{cases}$$
\item If $G=S_p$ is a noncyclic $p$-group, then $\boxi(\Pow(G))=1.$
\end{enumerate}
\end{theorem}

Because $h_3(S_2)$ is an integer or $-\infty$, the alternatives in Theorem~\ref{thm:main} exhaust all cases.

\noindent\textit{Overview of the methods used:}  To achieve the results, we have combined order-theoretic, geometric, topological, and group-theoretic arguments. We first replace the power graph by the comparability graph of the poset of cyclic subgroups, preserving boxicity. For a finite nilpotent group, the Sylow subgroup decomposition expresses this poset as a product of rooted tree posets, whose heights are determined by the exponents of the Sylow subgroups of the nilpotent group. Explicit interval constructions provide upper bounds for adding chain factors and replacing chains by rooted tree posets.

For the lower bounds, we use topological concepts, in particular a combination of the Van Kampen obstruction and Tancer’s nonrepresentability theorem with the \(2\)-Helly property of axis-parallel boxes. This yields a boxicity obstruction for comparability graphs of certain kinds of product posets. In particular, we determine the exact boxicity of the Boolean lattice and thereby settle the cyclic case. For more general products, we select layers that form an induced graph join and use the additivity of boxicity under joins. Group-theoretic arguments provide special subposet structures in the corresponding poset of cyclic subgroups, particularly \emph{terminal ternary forks} in each noncyclic odd Sylow factor.

\subsection{Organisation of the paper}
In \Cref{sec: preli}, we write the necessary preliminaries on posets, groups, and boxicity. In \Cref{sec: topological tool}, we establish the topological obstruction used for our lower bounds, after providing the necessary background. \Cref{secfr} develops bounds for comparability graphs of product posets and determines the exact values for mixed products of chains and stars. Further, we apply these tools to finite cyclic groups in \Cref{sec: cyclic main} and to nilpotent groups in \Cref{sec: pow nilpotent}. Appendix~\ref{appendix: proof lemma 4.18} gives an alternative combinatorial proof of the Boolean-lattice result, while Appendix~\ref{app: omitted proof} contains an omitted proof.

\section{Preliminaries}\label{sec: preli}

\subsection{Definitions on Posets}

A \emph{chain} is a poset where all elements are pairwise comparable. We denote a chain of length $e$ as $C_e=\{0<1<2<\dots<e\}$. 

\begin{definition}\label{def: rooted tree poset}
    A \emph{rooted tree poset} is a finite poset \((T,\le)\) with a unique minimum element \(\rho\), called its \emph{root}, such that, for every \(v\in T\), the set
\(\{u\in T:u<v\}\) is a chain. So, $\rho \le v$ for all $v \in T$. The \emph{depth} of a node \(v\) is defined by \(\depth(v)=|\{u \in T\ : \ u<v\}|\), i.e., the distance of \(v\) from the root $\rho$ in the cover graph of \(T\). 
\end{definition}

The \emph{height} of a rooted tree poset $T$ is the maximum length of the path from the root to a node, or the length of an induced longest chain in $T$. For two elements $u, v \in T$, if $u \le v$, then we say that $u$ is an \emph{ancestor} of $v$ and $v$ is a \emph{descendant} of $u$. A \emph{strict descendant} (or, \emph{ancestor}) of \(u\) is an element \(v\) satisfying \(u<v\) (or, $v<u$).  An \emph{immediate descendant} of \(u\) is an element \(v\) satisfying \(u<v \text{ and there is no $w$ such that } u<w<v\). The notion of \emph{immediate ancestor} of an element in the rooted tree poset $T$ should be clear. Maximal members (if any)
of a rooted tree are called \emph{leaves}, and maximal chains as induced subposets are called \emph{branches}. Note that the cover graph of a rooted tree poset $T$ is a tree. Every non-root element in $T$ has a unique immediate ancestor.

  We define a \emph{star poset} as a rooted tree poset when the cover graph is a star. In this paper, we use $\mathrm{S}$ to denote the four-element star poset consisting of a root element $\rho$ and three leaves $a_1,a_2,a_3$, where the relations in the poset are given by \( \rho<a_j, \ j=1,2,3\), with the leaves pairwise incomparable.

    A \emph{broom of height \(e\ge1\)} is a rooted tree poset with elements \(c_0,\ldots,c_{e-1},x_1,x_2,x_3\), whose order is determined by 
    $c_0<c_1<\cdots<c_{e-1}<x_i$ for $i\in\{1,2,3\}$ with \(x_1,x_2,x_3\) pairwise incomparable.

    For $k\ge2$, an element $u$ of a poset $P$ is said to have a \emph{$k$-ary fork} if $u$ has $k$ pairwise incomparable strict descendants in $P$. 
    
    A poset $P$ is called the \emph{product poset} of two posets $(P_1,\le_1)$ and $(P_2,\le_2)$, if $P=P_1\times P_2$ and $\forall (x_1,x_2),(y_1,y_2)\in P$, we have $(x_1,x_2)\le(y_1,y_2)\iff  x_1\leq _1 y_1 \text{ and }  x_2\leq _2 y_2.$ The \emph{comparability graph} $\Comp(P)$ of a poset $(P,\le)$ is a simple and undirected graph with vertex set $P$; two distinct vertices $x$ and $y$ are adjacent in $\Comp(P)$ if and only if $x$ and $y$ are comparable in $(P,\le)$, i.e., $x \le y$, or $y \le x$.

\subsection{Basics on Group Theory and Power Graphs} 

In this paper, we only consider finite groups. An interested reader can refer to any standard book (e.g., \cite{rotman2012introduction}) for the basic definitions and facts on group theory. The subgroup $H$ of a group $G$ is denoted by $H \le G$. The \textit{order of a group} $G$, denoted by $|G|$, is the number of elements in $G$. The \textit{order of an element} $g$ in $G$, denoted by $o(g)$, is the smallest positive integer $k$ such that $g^k=e$, where $e$ is the identity element of the group $G$. The \textit{exponent} of a finite group $G$, denoted by $\exp(G)$, is the least common multiple of the orders of all elements of $G$. 

The \emph{direct product} of two groups $G_1, G_2$, denoted by $G_1 \times G_2$, is the group on the elements $(g_1,g_2)$, $g_1 \in G_1, \ g_2 \in G_2$ and the group operation on $G_1 \times G_2$ is given as follows: $(g_1,g_2)(g_1',g_2')=(g_1g_1',g_2g_2')$ where the co-ordinate wise binary operations are the group operations of $G_1$ and $G_2$ respectively. A subset $T$ of elements of a group $G$ is said to \emph{generate a subgroup} $H\le G$ if $H$ is the set of all finite products of elements of $T$. The set $T$ is called \emph{a set of generators} of $H$, and we write $H=\ang{T}$. When $S$ is a singleton set $\{g\}$, then the subgroup $\ang{g}$ generated by $\{g\}$ is called the \emph{cyclic subgroup generated} by $g$. A cyclic group of order $n$ is isomorphic to $\mathbb{Z}_n$, the additive group of integers modulo $n$.

A group $G$ is called a \textit{$p$-group} if the order of $G$ is a power of a prime $p$. 
It is common to refer to a group as a $p$-group whenever its order is a power of some prime, where the symbol $p$ denotes an arbitrary prime. Note that the exponent $\exp(G)$ of a $p$-group $G$ is the maximum order of an element in $G$. If $p^m$ is the highest power of a prime $p$ such that $p^m$ divides $|G|$, then there is at least one subgroup of order $p^m$ in $G$, and such a subgroup is called a \textit{Sylow p-subgroup} of $G$. A finite group is a \textit{nilpotent group}\footnote{While the definition of nilpotent groups involves central series, this equivalent fact suffices for our purposes.} if and only if it is a direct product of its Sylow subgroups. Moreover, for each prime factor $p$ of $G$, the Sylow $p$-subgroup is unique in a finite nilpotent group $G$.

\begin{definition}\label{def:pow}
The \emph{power graph} $\Pow(G)$ of a finite group $G$ is a simple, undirected graph with vertex set $G$.  Two distinct elements $x$ and $y$ are adjacent in $\Pow(G)$ if and only if $x=y^k$ or $y=x^k$ for some
positive integer $k$.
\end{definition}

The following observation is straightforward from the definition.

\begin{observation}
    For two distinct elements $x, y\in G$, $\{x,y\}\text{ is an edge in }\Pow(G)$
 \text{ if and only if } $\langle x\rangle\le \langle y\rangle
 \text{ or }
 \langle y\rangle\le \langle x\rangle \text{ in } G$.
\end{observation}

A power graph is a comparability graph \cite{aalipour2017structure}. The relation between these two graphs, in terms of boxicity, is also supported by~\citep[Observation~3]{chandran2026boxicity}. The key idea behind this is a fundamental observation concerning the relationships among cyclic subgroups of a group, which provides a poset structure. Note that this poset structure is indeed the poset $\C(G)$ of cyclic subgroups of a group $G$, ordered by inclusion. The next lemma serves as a bridge between these two graphs.





\begin{lemma}\label{lem:reduction from pow to comp}
For every finite group $G$, $\boxi(\Pow(G))=\boxi(\Comp(\C(G)))$.

\end{lemma}




\subsection{Basics on Boxicity}

An equivalent formulation of boxicity was given by Roberts~\cite{Roberts} and will be used throughout the paper.

\begin{theorem}[\cite{Roberts}]\label{Roberts}
    A non-complete graph $G$ has boxicity at most $l>0$, if and only if it can be represented as the intersection of $l$ many interval graphs, i.e., there exist interval supergraphs $\{I_1, I_2, \ldots , I_{l}\}$ 
    of $G$ such that $\forall i\in [l]$, $V(I_i)=V(G)$ and $E(G)=E(I_1)\cap E(I_2)\cap \cdots \cap E(I_l)$. 
\end{theorem}

For vertex-disjoint graphs $\Gamma_1$ and $\Gamma_2$, their \emph{join} $\Gamma_1\vee \Gamma_2$ is obtained
from their disjoint union by adding every edge between $V(\Gamma_1)$ and $V(\Gamma_2)$.

\begin{lemma}[\cite{chandran2026survey}]\label{lem:join}
For vertex-disjoint graphs $\Gamma_1$ and $\Gamma_2$, $\boxi(\Gamma_1\vee\Gamma_2)=\boxi(\Gamma_1)+\boxi(\Gamma_2)$.

\end{lemma}

The following lemma is straightforward, and we state it without a proof.
\begin{lemma}[\cite{chandran2026survey}]
\label{lem:disjoint-union}
Let $d\ge1$. Let  $\Gamma_1$ and $\Gamma_2$ be vertex-disjoint graphs with boxicity at most $d$. Then $\Gamma_1\sqcup\Gamma_2$ has a $d$-box representation.
\end{lemma}

\section{The Topological Tool}\label{sec: topological tool}

\input{topological_tool}

\section{The Boxicity of Comparability Graph of Product of Posets}\label{secfr}

\input{comparability_graph_box_tool}

\input{some_particular_poset_structure}

\section{The Boxicity of Power Graphs of Nilpotent Groups}\label{sec: pow nilpotent}

We will now explore the boxicity of power graphs associated with nilpotent groups. In the first subsection \Cref{sec: structure for pow of nil}, we will examine the structure of power graphs that arise from finite nilpotent groups, specifically in relation to the cyclic subgroup poset. The next subsection (\Cref{sec: cyclic main}) focuses on cyclic groups, a specific subclass of finite nilpotent groups. The findings in this subsection primarily rely on \Cref{subsec:boolean-product} and previously established results from \cite{chandran2026boxicity}. In the subsequent sections, we will extend these results to encompass all finite nilpotent groups. We will first present an upper bound applicable to all finite nilpotent groups (\Cref{sec:upper}), and then we will proceed to determine the exact boxicity based on the structural characteristics of the nilpotent group (\Cref{sec:two}).

\subsection{Structure of Power Graphs of Finite Nilpotent Groups}\label{sec: structure for pow of nil}

In this section, we explore the structure of the power graph associated with a finite nilpotent group \( G \), focusing specifically on the poset \( \C(G) \) of cyclic subgroups of \( G \). Our objective is to investigate the boxicity of these graphs, and \Cref{lem:reduction from pow to comp} suggests that a thorough analysis of the poset structure suffices for our purposes. Although \Cref{eq: nilpotent} provides the final expression, it is essential to first write the preliminary lemma. This lemma follows almost immediately as a corollary from the discussion of \emph{subgroup lattice} of direct products of groups in \cite[Lemma~1.6.4]{schmidt1994subgroup}. By doing their arguments for the cyclic subgroups, we get \Cref{lem:sylow-product}. A more general statement is mentioned in \cite[Section~2]{garonzi2018number} in terms of the number of cyclic subgroups of the direct product of two groups with coprime order. 

\begin{lemma}\label{lem:sylow-product}
Let $G$ be a finite nilpotent group and $G=\prod_{i}S_{p_i}$, where $S_{p_i}$ denotes the Sylow $p_i$-subgroup of $G$. Then $\C(G)\cong\prod_i\C(S_{p_i})$ as the product of posets.
\end{lemma}

The following lemma collects standard facts about the cyclic-subgroup poset of a finite \(p\)-group. The tree property is explicitly noted in the proof of Proposition~4.1 of Leifman and Muzychuk~\cite{LeifmanMuzychuk2005}; it also follows from the edge-counting formula in Lemma~2.2 of Tărnăuceanu~\cite{Tarnauceanu2023}, as illustrated in the proof of his Theorem~1.1. Cameron, Reddy and Sharma~\cite[Theorems~2.6 and~2.4(2)]{CameronReddySharma} give tree and path characterisations for \emph{cyclic subgroup graphs}, which yield the corresponding tree and chain assertions here. We include a self-contained proof for completeness and to record the relation between the height of the poset and the exponent of the group explicitly.

\begin{lemma}\label{lem:p-tree}
Let $S_p$ be a finite $p$-group of exponent $p^e$.
\begin{enumerate}[label=\textup{(\alph*)}]
\item $\C(S_p)$ is a rooted tree poset of height $e$.
\item $\C(S_p)$ is a chain if and only if $S_p$ is cyclic.
\end{enumerate}
\end{lemma}

\begin{proof}
(a) Let $U=\langle u\rangle$ be cyclic of order $p^k$, where $k\ge1$.  A cyclic group of order
$p^k$ has exactly one subgroup of order $p^{k-1}$, namely
$
 \langle u^p\rangle.
$
Therefore, every non-root node of $\C(S_p)$ has a unique element strictly below it in the cover graph.  Repeatedly
taking the lower cover reaches the trivial subgroup, so the cover graph is
a rooted tree.

Because $S_p$ has exponent $p^e$, it contains an element of order $p^e$ and no
element of larger order.  The maximum root-to-node distance is therefore
$e$.

(b) If $S_p$ is cyclic, it has exactly one subgroup of each order $1,p,\ldots,p^e$,
so $\C(S_p)$ is a chain.  Conversely, suppose $\C(S_p)$ is a chain.  Let $U$ be
its largest member.  For every $x\in S_p$, the cyclic subgroup $\langle x\rangle$
is comparable with and hence contained in $U$.  Thus, every element of $S_p$
lies in $U$, so $S_p=U$ is cyclic.
\end{proof}

Combining Lemmas~\ref{lem:reduction from pow to comp}, \ref{lem:sylow-product}, and
\ref{lem:p-tree}, we obtain the following formulation, when $G = \prod_i S_{p_i}$ is a finite nilpotent group:
\begin{equation}\label{eq:central-reformulation}
 \boxi(\Pow(G))
 =\boxi\!\left(\Comp\!\left(\prod_i T_i\right)\right),
 \qquad T_i=\C(S_{p_i}),
\end{equation}
where $T_i$ is a rooted tree poset of height $e_i$ (where $\exp(S_{p_i})=p_i^{e_i}$) and is a chain exactly when
$S_{p_i}$ is cyclic. In particular, we have the following for the power graph of the cyclic group $\Z_n$, where $n=\prod_i p_i^{e_i}$:

\begin{equation}\label{eq:cyclic central-reformulation}
 \boxi(\Pow(\Z_n))
 =\boxi\!\left(\Comp\!\left(\prod_i C_i\right)\right),
\end{equation}
where $C_i$ is a chain of length $e_i$.

 \subsection{The Boxicity for Power Graphs of Cyclic Groups}\label{sec: cyclic main}

For a cyclic group $\Z_n$ of order $n=\prod_i p_i^{e_i}$, \Cref{eq:central-reformulation} gives the following: $ \boxi(\Pow(\Z_n))
 =\boxi\!\left(\Comp\!\left(\prod_iC_{e_i}\right)\right)$, where $C_{e_i}$ denotes the chain poset of length $e_i$. 
Recall that Chandran and Ghosh \cite{chandran2026boxicity} established the following upper bound, which is also stated in (a) of \Cref{prevthm}. 

\begin{lemma}\cite[Cor.~1]{chandran2026boxicity}\label{cyclic upper bound known}
    Let $n=\prod_i^d p_i^{e_i}$. Then $\boxi(\Pow(\Z_n)) \le E-M$, where $E=\sum_{i}e_i$ and $M=\max_i e_i$.
\end{lemma}


 Chandran and Ghosh~\cite{chandran2026boxicity} showed the following lemma.

  \begin{lemma}[\cite{chandran2026boxicity}]\label{lem: cyclic lower bound general formula}
    Let $1\leq e_1 \leq e_2 \leq \dots \leq e_d$ be $d$ natural numbers, where $d \geq 2$. Let $n=p_1^{e_1}p_2^{e_2}\dots p_d^{e_d}$, where $p_1,p_2,\ldots,p_d$ are pairwise distinct primes. Then $\boxi(\Pow(\Z_n)) \geq e_1 f(d) + (e_2-e_1) f(d-1) + (e_3-e_2) f(d-2) + \dots + (e_{d-1}-e_{d-2}) f(2)$.
\end{lemma}

 In fact, parts (b) and (c) of Theorem~\ref{prevthm} are a consequence of Lemma~\ref{lem: cyclic lower bound general formula} and the fact that for $r\ge 2$, $f(r)\geq \left\lceil \frac{r}{2}\right\rceil$. Consequently, an improvement of the lower bound on $f(r)$ gives an improved lower bound for $\boxi(\Pow(\Z_n))$ for free. 
Substituting the optimal value of $f(r)$, which is $f(r)=r-1$ (obtained in \Cref{lem:boolean}, and also in \Cref{lem1}) for $r=d,d-1,\dots,2$, in the above formula of \Cref{lem: cyclic lower bound general formula}, we get the following lower bound. Note that this lower bound matches the upper bound proved by Chandran and Ghosh \cite{chandran2026boxicity}, stated in \Cref{cyclic upper bound known}.








\begin{lemma}\label{thm:product-chains}
Let $e_1,\ldots,e_d\ge1$. Then $\boxi(\Comp(\prod_{i=1}^dC_{e_i})) \ge E-M.$
\end{lemma}

Now we are ready to prove our first main theorem, \Cref{thm: cyclic}.

\noindent\textit{Proof of \Cref{thm: cyclic}}: Combining \Cref{thm:product-chains}, \Cref{eq:central-reformulation} and \Cref{cyclic upper bound known}, we have the following: if $n=\prod_{i=1}^dp_i^{e_i}$, then
$\boxi(\Pow(\Z_n))=\boxi(\Comp(\prod_iC_{e_i}))
 =E-M$. \hfill $\square$

\subsection{The Universal Upper Bound for Arbitrary Nilpotent Groups}\label{sec:upper}

We now extend the cyclic upper bound to arbitrary finite nilpotent groups. The proof is relatively straightforward and can be easily derived from the lemmas and tools we have previously discussed.

\begin{lemma}\label{thm:universal-upper}
Let $G=S_{p_1}\times S_{p_2}\times\ldots \times S_{p_k}$ be a finite nilpotent group. Then $\boxi(\Pow(G))\le \boxi(\Pow(\Z_{\exp(G)}))+s(G)$, where $s(G)$ is the number of non-cyclic Sylow subgroups of $G$.
\end{lemma}

\begin{proof}
Suppose $\exp(S_{p_i})=p^{e_i}$. Note that $\exp(G)=\prod_i p_i^{e_i}$ (by using Lemma~\ref{lem:sylow-product}). Due to \Cref{eq:central-reformulation}, it is enough to prove the upper bound of $\boxi(\Comp(\prod_{i}T_{i}))$, where $T_{i}$ denotes the rooted tree poset arising from the Sylow $p_i$-subgroup $S_{p_i}$. Since $\exp(S_{p_i})=p^{e_i}$, height of the rooted tree poset $T_i$ is $e_i$. 
Let $S(G)=\{i \in [k] : \text{ $S_{p_i}$ is non-cyclic }\}$. Then $s(G)=|S(G)|$. Hence, from \Cref{eq:central-reformulation}, it is clear that for $i \notin S(G)$, the rooted tree poset $T_{i}$ is indeed a chain of height/length $e_i$. 
\[
\begin{aligned}
\boxi(\Comp(\prod_{i}T_{i})) & = \boxi(\Comp(\prod_{i\in S(G)}T_{i}\times\prod_{i \notin S(G)}C_{e_i}))\\
 & \le \boxi(\Comp(\prod_i C_{e_i})) + s(G)&&\text{(Lemma~\ref{lem:tree-refinement})}\\ 
 & = \boxi(\Pow(\Z_{\exp(G)})) + s(G).&&\text{(\Cref{eq:cyclic central-reformulation})}
 \end{aligned}
 \]

\end{proof}

\subsection{Lower Bound Due to Odd Order Non-Cyclic Sylow Subgroups}\label{sec:forks}

The previous upper bound of $\boxi(\Pow(G))$ for an arbitrary nilpotent group $G$ treats each rooted tree poset $T_{i}$, which arises from a non-cyclic Sylow subgroup $S_{p_i}$ of $G$, uniformly. For the lower bound, we observe that the structure of a rooted tree poset necessitates the introduction of an additional dimension, coordinate, or interval. We also show that for odd primes, there is sufficient branching at the terminal level of the rooted tree posets associated with the corresponding Sylow subgroups.

\input{cyclic_overgroups_odd_p_group}

\input{lower_bound_proof_for_odd_p}


\subsection{Boxicity Due to Non-cyclic Sylow \texorpdfstring{$2$}{2}-subgroup}
\label{sec:two}

Recall that for $k\ge2$, an element $u$ of a poset $P$ is said to have a $k$-ary fork if $u$ has $k$ pairwise incomparable strict descendants in $P$. At an odd prime, Lemma~\ref{lem: terminal fork for odd prime} supplies a ternary fork one
level below the exponent.  This fails at $p=2$.  For example, the cyclic
subgroup poset (also a rooted tree poset) of a dihedral $2$-group has one long rotation chain, while its
reflection subgroups branch only from the root.  We therefore need to record
the depth of branching rather than merely whether the rooted tree poset is a chain.

The next lemma shows that a deep binary fork in the cyclic-subgroup rooted tree poset of a $2$-group always creates a ternary fork at the same depth or one level lower.
This special fact is false for arbitrary rooted tree posets; its proof uses group
multiplication.

\begin{lemma}
\label{lem:h2-h3}
For every noncyclic finite $2$-group $G$, $h_2(G)\le h_3(G)+1.$
\end{lemma}

\begin{proof}

Let $d=h_2(G)$.

First, suppose $d=0$. Then every element of $G$ lies in a maximal cyclic subgroup.  A group cannot be
the union of two proper subgroups (see \citep{Bhargava} for details). 
Since $G$ is noncyclic, it follows that $G$ has at least three maximal cyclic subgroups.  These
are pairwise incomparable, and all contain the trivial subgroup.  Hence
$h_3(G)\ge0$, and the lemma follows in this case.

Now suppose $d\ge1$. Choose a cyclic subgroup \(U=\langle u\rangle\) of order \(2^d\)
contained in two incomparable cyclic subgroups \(A\) and \(B\).
Let \(X=\langle x\rangle\) and \(Y=\langle y\rangle\) be their
respective unique subgroups of order \(2^{d+1}\). Note that both contain \(U\).
If \(X=Y\), this common subgroup is strictly contained in both
\(A\) and \(B\), since \(A\) and \(B\) are incomparable.
It would therefore support a binary fork at depth \(d+1\),
contrary to \(d=h_2(G)\).
Thus \(X\ne Y\). Since \(U\le X\cap Y\) and \(U\) has index \(2\)
in \(X\), while \(X\) and \(Y\) have the same order, it follows
that \(X\cap Y=U\). 



Note that without loss of generality, we can put \(u=x^2\). Since \(y^2=u^s\) for some odd integer \(s\), replacing \(y\) by \(y^t\), where \(st\equiv1\pmod{2^d}\), gives
\(x^2=y^2=u.\) The integer \(t\) is odd, so this replacement preserves \(\langle y\rangle=Y\). We define the subgroup \(H=\langle x,y\rangle\). Both generators centralize \(u\), and therefore \(U\le Z(H)\).

Write \(X^y=y^{-1}Xy\). If \(X^y\ne X\), then \(X^y\) contains \(U\) and differs from \(Y\): equality would imply \(X=Y^{y^{-1}}=Y\). Thus \(X,X^y,Y\) are three distinct cyclic overgroups of \(U\) of the same order, yielding \(h_3(G)\ge d\).

Suppose now that \(X^y=X\). Then \(y^{-1}xy=x^r\) for some odd integer \(r\), so
\([x,y]=x^{-1}y^{-1}xy=x^{r-1}\in\langle x^2\rangle=U.\)
Write \([x,y]=u^k\). Since \(U\le Z(H)\), the commutator is central, and
\[
\begin{aligned}
(xy)^2&=x^2y^2[x,y]^{-1}=u^{2-k},\\
(xy^{-1})^2&=x^2y^{-2}[x,y]=u^k.
\end{aligned}
\tag{*}
\]
Moreover, \(H/U\cong C_2\times C_2\), because it is generated by the distinct nonidentity commuting involutions \(xU\) and \(yU\).

If \(k\) is odd, take \(z=xy^{-1}\). Then \(\langle z^2\rangle=U\). If \(k\) is even, put \(W=\langle u^2\rangle\). When \(d=1\), take \(z=xy\); equation \((*)\) gives \(z^2=1\), so \(\langle z^2\rangle=W\). When \(d\ge2\), exactly one of \(k\) and \(2-k\) is congruent to \(2\pmod4\). Choose the corresponding \(z\in\{xy^{-1},xy\}\); then \((*)\) gives \(\langle z^2\rangle=W\).

Thus, in every case, \(Z=\langle z\rangle\) contains a subgroup \(W_0=\langle z^2\rangle\le U,
\ \operatorname{depth}(W_0)\ge d-1.\)
Under the quotient map \(H\to H/U\), the images of \(X,Y,Z\) are the three distinct subgroups of order \(2\), since \(zU=xyU\). Subgroup inclusion is preserved by homomorphisms, so \(X,Y,Z\) are pairwise incomparable. They all contain \(W_0\), and hence form a ternary fork at depth at least \(d-1\). Therefore, \(
h_3(G)\ge d-1=h_2(G)-1.
\)

\end{proof}

\begin{remark}\label{rem:fork-down}
If a node $u$ at depth $d$ lies below $k$ incomparable descendants, then
every ancestor of $u$ also lies below those same descendants.  Consequently,
a fork known at depth at least $r$ may always be used at any smaller depth by
replacing its core with the appropriate ancestor.  We will use this when
$a>m$ in \Cref{thm:noncyclic-two-classification}.
\end{remark}

\begin{corollary}
\label{cor:branching-consequences}
Let $q\ge1$.  Then
\(
 h_3(S_2)\le q-2\implies h_2(S_2)\leq q-1.
\)

\end{corollary}
\begin{proof}
If not, Lemma~\ref{lem:h2-h3} would produce a ternary fork at
depth at least $d-1\ge q-1$, contradicting $h_3(S_2)\le q-2$.
\end{proof}

\begin{proposition}
\label{prop:two-group-absorption}
Let $S_2$ be a finite $2$-group of exponent $2^a$, let $m\ge1$, and put
$q=\min\{a,m\}$.  If
\(
 h_3(S_2)\le q-2,
\)
then
\begin{equation}\label{eq:two-group-absorption}
 \boxi\bigl(\Comp(\C(S_2)\times C_m)\bigr)=q.
\end{equation}

\end{proposition}

\begin{proof}
By Corollary~\ref{cor:branching-consequences}, the rooted tree poset
$T_2=\C(S_2)$ satisfies $h_3(T_2)\leq q-2$ and $h_2(T_2)\leq q-1$. By
Lemma~\ref{lem:low-branching-absorption}, the stated equality follows.
\end{proof}

We now prove the even-order classification.  Assume first that the odd part
of $G$ is nontrivial, and write
\(
 a=e_2,
 \
 m=\max_{p\text{ odd}}e_p,
 \
 q=\min\{a,m\},
 \
 E_{\mathrm{odd}}=\sum_{p\text{ odd}}e_p.
\)
Let
\(
 t=\bigl|\{p\text{ odd}:S_p\text{ is noncyclic}\}\bigr|,
 \
 B=a+E_{\mathrm{odd}}-\max\{a,m\},
 \
 L=B+t.
\)

\begin{lemma}
\label{thm:noncyclic-two-classification}
Let $G$ be a finite nilpotent group whose Sylow $2$-subgroup $S_2$ is
noncyclic and whose odd part is nontrivial.  With the notation above,
\begin{equation}\label{eq:h3-only-classification}
 \boxi(\Pow(G))=
 \begin{cases}
 L+1,&h_3(S_2)\ge q-1,\\[1mm]
 L,&h_3(S_2)\le q-2.
 \end{cases}
\end{equation}
\end{lemma}

\begin{proof}
For each prime \(p\mid o(G)\), choose an induced subposet
\(P_p\) of \(\mathcal C(S_p)\) as follows. 

For every noncyclic odd Sylow subgroup $S_p$,
Lemma~\ref{lem: terminal fork for odd prime} supplies a cyclic subgroup at depth
$e_p-1$ contained in three incomparable cyclic subgroups at depth $e_p$. This induces a broom\footnote{Recall that, a broom of height $e$ is a rooted tree poset with elements \(c_0,\ldots,c_{e-1},x_1,x_2,x_3\), whose order is determined by 
    $c_0<c_1<\cdots<c_{e-1}<x_i$ for $i\in\{1,2,3\}$ with \(x_1,x_2,x_3\) pairwise incomparable.}. If \(h_3(S_2)\ge q-1\), choose, using Remark~\ref{rem:fork-down}, a node \(U\)
at depth \(q-1\) with three pairwise incomparable strict descendants. Let \(P_2\) be the induced subposet consisting of the root-to-\(U\) chain and these three descendants; thus \(P_2\) is a broom of height \(q\). If \(h_3(S_2)\le q-2\), let \(P_2\) be a root-to-leaf chain of height \(a\).

On the other hand, if $S_p$ is cyclic,
let $P_p$ be its unique chain of height $e_p$.  The product
\(
 P_2\times\prod_{p\text{ odd}}P_p
\)
is an induced subposet of $\C(G)$.  Apply Lemma~\ref{lem:scheduled-forks} with \(R\) consisting of the \(t\) noncyclic odd
factors, together with the \(2\)-factor when \(h_3(S_2)\ge q-1\).
Observe the sum of the factor heights minus their maximum. When \(h_3(S_2)\le q-2\), it is
\(a+E_{\mathrm{odd}}-\max\{a,m\}=B\),
and when \(h_3(S_2)\ge q-1\), it is \(q+E_{\mathrm{odd}}-\max\{q,m\}=q+E_{\mathrm{odd}}-m=B\) since
\(q=\min\{a,m\}\) and $\max\{a,m\}+\min\{a,m\}=a+m$. 

Therefore, we have
\begin{equation}\label{eq:two-value-lower}
 \boxi(\Pow(G))\ge \begin{cases}
     B+t=L & \text{ if } h_3(S_2)\leq q-2\\
     B+t+1=L+1 & \text{ if } h_3(S_2)\geq q-1
 \end{cases} 
\end{equation}
Since exactly $t+1$ Sylow subgroups are noncyclic, the universal upper bound
in \Cref{thm:universal-upper} gives
\begin{equation}\label{eq:two-value-upper}
 \boxi(\Pow(G))\le B+t+1=L+1.
\end{equation}
Thus, only the two values $L$ and $L+1$ are possible.

Let $h_3(S_2)\le q-2$. Choose an odd prime $p_0$ with $e_{p_0}=m$. Proposition~\ref{prop:two-group-absorption} gives
\begin{equation}\label{eq:T-times-longest-odd-chain}
 \boxi\bigl(\Comp(\C (S_2)\times C_m)\bigr)=q.
\end{equation}
Therefore, we have 
\[
\begin{aligned}
\boxi(\Pow(G))&=\operatorname{box}\!\left(
  \operatorname{Comp}\!\left(\C(S_2)\times\prod_{p\text{ odd}}\C(S_p)\right)
\right)\\
&\le
\operatorname{box}\!\left(
  \operatorname{Comp}\!\left(\C(S_2)\times\prod_{p\text{ odd}}C_{e_p}\right)
\right)+t&&\text{(Lemma~\ref{lem:tree-refinement})}\\
&\le
\operatorname{box}\!\left(
  \operatorname{Comp}(\C(S_2)\times C_m)
\right)+E_{\mathrm{odd}}-m+t&&\text{(Lemma~\ref{lem:add-chain})}\\
&=q+E_{\mathrm{odd}}-m+t&&\text{(\Cref{eq:T-times-longest-odd-chain})}\\
&=L.
\end{aligned}
\]
\end{proof}

\noindent\textit{Proof of \Cref{thm:main}} 
\begin{itemize}
    \item[(i)] This follows from \Cref{thm:universal-upper} and \Cref{cor:odd-cyclic2}.
    \item[(ii)] This is \Cref{thm:noncyclic-two-classification}.
    \item[(iii)] This follows from \Cref{lem:p-tree} and \Cref{lem:dfs}. Note that a noncyclic $p$-group corresponds to a cyclic subgroup poset, which is non-chain, and hence boxicity is at least $1$. \hfill $\square$
\end{itemize}

\textbf{Declaration of AI use:} Our preliminary investigation of the boxicity of power graphs of cyclic groups led us to its complete solution via Lemma~\ref{lem1}, without any use of AI tools. An upper bound for the general nilpotent case was also similarly derived. As a result of playing with small cases, we observed that for odd primes $p$ and $q$, the reduced power graph of the group $\Z_{pq}\times \Z_{pq}$ contains an isomorphic copy of the fully subdivided $K_{3,3}$ as an induced subgraph. It is known that the boxicity of the full subdivision of any non-planar graph is at least 3\cite{6767831,EHRLICH19768}. This observation led us to believe that there might be an underlying topological obstruction affecting the representability of these graphs, even for more general nilpotent groups. This curiosity led us to look for such an obstruction, for which we used ChatGPT 5.6 Sol. The LLM was able to find exactly the kind of structure that we were looking for. The text has been carefully vetted for mistakes. None of the references is generated by the LLM. We take full responsibility for the contents of this paper.

\bibliographystyle{plainnat}
\bibliography{reference}

\appendix

\section{Alternate proof of \Cref{lem:boolean}}\label{appendix: proof lemma 4.18}

The alternate proof is via the proof of \Cref{lem1}. This technique uses the notion of a \emph{divisor graph}, so we need some additional definitions and facts. We first set the background for that.

\begin{definition}
    Define $g\colon \Z^+\rightarrow \Z^+\cup \{0\}$ as $g(x)\colon=$ the number of distinct prime divisors in the prime factorisation of $x$.
\end{definition} 

\begin{definition}
    For a positive integer $n$ let $D(n)$ be the graph with vertex set equal to the set of all non-trivial proper divisors of $n$, such that two vertices are adjacent in $D(n)$ if one divides the other. $D(n)$ is called the divisor graph of $n$. 
\end{definition}

Let $n$ be a square-free positive integer. Suppose $g(n)=r$. It was observed in~\cite{chandran2026boxicity} and also in \Cref{subsec:boolean-product} that $\boxi(D(n))=\boxi(TC(H_r))$. Therefore, it is enough to compute the exact boxicity of $D(n)$.

\begin{definition}
   For a positive integer $n$, define $\mathcal{D}(n)$ to be the collection of all graphs $\Gamma$ such that $V(\Gamma)=V(D(n))$ and $E(D(n))\subseteq E(\Gamma)$ such that $E(\Gamma)$ does not contain the edges $\{xy\colon x,y\in V(\Gamma) \text{ and } y=\frac{n}{x}\}$.
   Define $$\tilde{\mathcal{D}}(r)\colon =\bigcup_{\substack{g(n)=r\\n\text{ is square free}}}\mathcal{D}(n).$$
   
\end{definition}

The following two statements are used in the proof of \Cref{lem1}.
\begin{observation}\label{6graph}
    Let $\G$ be a graph having $V(\G)=\{v_1, v_2, \ldots, v_6\}$ such that:
    \begin{itemize}
        \item $v_1,v_2,v_3,v_4,v_5, v_6$ forms a cycle in $\G$, and
        \item $v_1v_4,v_2v_5,v_3v_6\notin E(\G)$.
    \end{itemize}
    Then $\boxi(\G)>1$.
\end{observation}
The observation follows by noting that graphs defined as above always contain $C_4,C_5,C_6$ or the Hajós graph as an induced subgraph, all of which are forbidden induced subgraphs for interval graphs. The following more general theorem is by Cozzens and Roberts~\cite{cozzens1983computing}. 
\begin{theorem}[\cite{cozzens1983computing}]\label{obs2}
   Let $\G$ be a graph and $p\geq 1$. Suppose that $V(\G)$ contains two disjoint sets of vertices $S_1=\{a_1, a_2, \ldots, a_{2p-1}\}$ and $S_2 = \{b_1, b_2,\ldots , b_{2p-1}$\}, and suppose that the only edges between $S_1$ and $S_2$ in $\overline{\G}$ are $\{a_ib_i\colon i\in [2p-1]\}$. Then $\boxi(\G)\geq p$. 
\end{theorem}

\begin{lemma}\label{lem1}
    The boxicity of any graph in $\tilde{\mathcal{D}}(r)$ is at least $r-1$. 
\end{lemma}
\begin{proof}
    We will proceed by induction on the number of primes. For $r\leq 2$, it is obvious. For $r=3$, the statement follows from Observation \ref{6graph}. So assume $r\geq 4$. 
    
    Base case: Suppose $r=4$. Suppose $n$ is a positive integer and suppose $n=p_1p_2p_3p_4$ is its prime factorisation. Suppose $\Gamma\in {\mathcal{D}}(n)$. Let, for the sake of contradiction, $\boxi(\Gamma)\leq 2$ and $I_1,I_2$ be an optimal box representation of $\Gamma$. Take $S=\{\frac{n}{p_4}, \frac{n}{p_3},\frac{n}{p_2},\frac{n}{p_1},p_4, p_3,p_2,p_1\}$. The graph induced by $S$ in $\Gamma$ is the graph as described in Theorem~\ref{obs2}. Therefore, the interval supergraphs $I_1$ and $I_2$ each contain exactly two of the pairs $\{\frac{n}{p_4}, p_4\},\{\frac{n}{p_3}, p_3\},\{\frac{n}{p_2}, p_2\},\{\frac{n}{p_1}, p_1\} $ because otherwise, there is a subgraph of the form described in Observation~\ref{6graph}. Suppose without loss of generality, $I_1$ contains $\{\frac{n}{p_4}, p_4\},\{\frac{n}{p_3}, p_3\}$ as non-edges. Then the graph induced by $\{p_1p_2,p_3p_4,\frac{n}{p_4}, p_4,\frac{n}{p_3}, p_3\}$ in $I_1$ is the graph of the form described as in Observation~\ref{6graph} unless there is an edge between $p_1p_2,p_3p_4$. Similarly, in $I_2$, the graph induced by $\{p_1p_2,p_3p_4,\frac{n}{p_2},\frac{n}{p_1}, p_1,p_2\}$ is the graph of the form described in Observation~\ref{6graph} unless there is an edge between $p_1p_2,p_3p_4$. But this forces $p_1p_2,p_3p_4$ to be an edge in $E(I_1)\cap E(I_2)=E(\Gamma)$. We have a contradiction.

    Now suppose the statement is true for $\tilde{\mathcal{D}}(r)$ for $r\leq l-1$. We need to show that $\tilde{\mathcal{D}}(l)$ has boxicity at least $l-1$. Suppose $n$ is a positive integer and suppose $n=\prod_{i\in [l]}p_i$ is its prime factorisation. Let $\Gamma\in \mathcal{D}(n)$. 
    
    Suppose $\boxi(\Gamma)=k$ and assume that $I_1, I_2, \ldots , I_k$ is an optimal box representation of $\Gamma$. For $i\in [l]$, define $S_{i}\colon=\{\frac{n}{p_i},p_i\}$ and let $\mathcal{S}=\{S_1, S_2, \ldots , S_l\}$.  If each $I_j$ contains at most one $S_i$ as an independent set, then $k\geq l\geq l-1$ and we are done. Therefore, assume that there exists $j'\in [k],i_1,i_2\in [l]$ such that $I_{j'}$ contains $S_{i_1}$ and $S_{i_2}$ as independent sets.  Without loss of generality, we can assume that $i_1=1,i_2=2$ and $j'=1$. This means that the set $\{\frac{n}{p_1p_2},p_1p_2,\frac{n}{p_1},\frac{n}{p_2},p_2, p_1\}$ induces in $I_1$ the graph described as in Observation~\ref{6graph} unless there is an edge between $\frac{n}{p_1p_2}$ and $p_1p_2$. By a similar argument, we can conclude that all the edges of the form $\{xy\colon y=\frac{n}{x} \text{ and }p_1p_2 \text{ divides } y \}$ are present in $I_1$. 
    
    \begin{claim}
        The graph $\bigcap_{j\in [k]\setminus \{1\}} I_j$ contains a graph from $\tilde{\mathcal{D}}(l-1)$ as an induced subgraph.
    \end{claim}
    \begin{claimproof}
      Let $\tilde{\Gamma}=\bigcap_{j\in [k]\setminus \{1\}} I_j$. Define $T_1=\{x\in V(\tilde{\Gamma})\colon  p_1p_2 \text{ divides }x\}$ and $T_2=\{x\in V(\tilde{\Gamma})\colon \text{neither } p_1 \text{ nor }p_2 \text{ divides }x\}$. Fix a prime $p_0$ which is not a divisor of $n$. Define $n'=\frac{p_0n}{p_1p_2}$. Construct $\tilde{\Gamma}'$ by removing all vertices in $V(\tilde{\Gamma})\setminus (T_1\cup T_2)$ and relabeling all vertices $x\in T_1$ as $\frac{p_0x}{p_1p_2}$. We will show that the graph $\tilde{\Gamma}'\in \tilde{\mathcal{D}}(l-1)$. It is clear that $V(\tilde{\Gamma}')=V(D(n'))$ and $E(\tilde{\Gamma}')\supseteq E(D(n'))$. Suppose $x,y\in V(\tilde{\Gamma}')$ such that $y=\frac{n'}{x}$. By definition, $p_0$ divides exactly one of $y$ or $x$. Without loss of generality, assume that $p_0$ divides $x$. Then $\frac{p_1p_2x}{p_0}\in T_1$ and $y\in T_2$. Then $\frac{p_1p_2x}{p_0}y\in E(I_1)$ but $\frac{p_1p_2x}{p_0}y\notin E(\Gamma)$. Therefore, $\frac{p_1p_2x}{p_0}y\notin E(\tilde{\Gamma})$ and it follows that $xy\notin E(\tilde{\Gamma}')$. Therefore, $\tilde{\Gamma}'\in \mathcal{D}(n')\subset \tilde{\mathcal{D}}(l-1)$.
    \end{claimproof}
    
    By the induction hypothesis and by the above claim, $k-1\geq l-2$ and we are done. 
\end{proof}

\section{Omitted Proof}\label{app: omitted proof}

\subsection{Proof of \Cref{obs: H_r is an induced subgraph}}\label{app: proof of H_r induced subgraph}

\begin{proof}
First, we label the vertices of $\dd$ by ordered pairs
\((i,j),\ i\in[r],\ j\in\{1,2,3\}\), where the index $i$ represents one of the $r$ copies of $D_3$ and $j$ represents which vertex to choose from that copy of $D_3=\{\emptyset,\{1\},\{2\},\{3\}\}$. 

For \(\mathbf z=(z_1,\ldots,z_r)\in \mathrm{S}^r\setminus\{\rho\},\) we define the mapping $\Phi : \mathrm{S}^r \setminus \{\rho\} \rightarrow \Pt(\dd)$ as follows: $ \Phi(\mathbf z)=\{(i,j):z_i=a_j\}$. 
Since each coordinate $z_i$ is either the root or one of the three leaves,
$\Phi(\mathbf z)$ selects at most one vertex from each copy of $\dd$. So, $\Phi(\mathbf z)$ is a face of $\dd$. Conversely, a non-empty face $\sigma\in \dd$ uniquely determines a tuple $\mathbf{z}$: use $a_j$ in coordinate $i$ when $(i,j)\in\sigma$, and use $\rho$ when the face makes no choice from the $i$-th copy.  Hence $\Phi$ is a bijection.

Furthermore, \(\mathbf z\le \mathbf z'\text{ in }\mathrm{S}^r \Longleftrightarrow  \Phi(\mathbf z)\subseteq\Phi(\mathbf z').\) Indeed, $\mathbf z<\mathbf z'$ holds exactly when for certain coordinates we have the relations $\rho<a_1,\rho<a_2$ or $\rho<a_3$ and $\Phi(\mathbf z)\subsetneq \Phi(\mathbf z')$ occurs exactly when certain factors of $D_3$ in $D_3^{*r}$ do not contribute to $\Phi(\mathbf z)$. These two situations occur one-to-one: $\rho$ appears in a coordinate of $\mathbf z$ if and only if that particular factor of $D_3$ in $D_3^{*r}$ does not contribute to $\Phi(\mathbf z)$. 
Therefore, $\Phi$ is an order isomorphism. 
The all-root tuple \( \boldsymbol\rho=(\rho,\ldots,\rho)\) in $\mathrm{S}^r$
corresponds to the empty face of $\dd$.  Consequently,
$\Pt(\dd) \cong \mathrm{S}^r\setminus\{\boldsymbol\rho\}.$ It follows that
$\G_r=\Comp(\Pt(\dd))$ is the subgraph of $\Comp(\mathrm{S}^r)$ induced by $\mathrm{S}^r\setminus\{\boldsymbol\rho\}.$ 
\end{proof}

\end{document}

%% file: topological_tool.tex
A `topological obstruction' will be crucial in our lower bound proofs. This subsection discusses the tool and provides the necessary definitions to build the background for the reader.

\subsection{Simplicial Complex, Nerve, and the Topological Obstruction}
\label{subsec:topological-obstruction}

In this subsection, we discuss the topological terminologies used in our paper. 
An interested reader can refer to any standard textbook on topology, such as \cite{matouvsek2003using}, for more detailed information.

\begin{definition}
Let $V$ be a finite set.  A \emph{simplicial complex} on $V$ is a family
$\kk$ of subsets of $V$ such that the following properties hold: (i) $\emptyset \in \kk$; and (ii) whenever $\sigma\in \kk$ and $\tau\subseteq\sigma$, one has
$\tau\in \kk$. The members of $K$ are called \emph{faces} or \emph{simplices}.
\end{definition}
By definition, a face $\sigma \in \kk$ containing $s$ vertices
has \emph{dimension} $s-1$, and $\dim \kk=\max_{\sigma\in \kk}\dim \sigma$. Thus, a zero-dimensional complex consists of isolated vertices together with the empty face. The \emph{face poset} or \emph{nonempty face poset} of a simplicial complex $\kk$ is
\(\Pt(\kk)=\kk\setminus\{\emptyset\}\), ordered by set inclusion. The \emph{order complex} $\Delta(P)$ of a finite poset $P$ is the simplicial complex whose vertex set is $P$ and whose faces are the finite chains of $P$.  

\begin{definition}
The \emph{barycentric subdivision} of a simplicial complex $\kk$, denoted by $\sd \kk$, is defined
abstractly as follows: $\sd \kk=\Delta\bigl(\Pt(\kk)\bigr)$.
Consequently, the following hold:
\begin{itemize}
\item a vertex of $\sd \kk$ is a nonempty face of $\kk$; and
\item a collection of such vertices is a face of $\sd \kk$ precisely when the corresponding faces of $\kk$ form an inclusion chain.
\end{itemize}
\end{definition}


\begin{definition}
Let $\mathcal A=\{A_i\}_{i\in I}$ be a family of sets indexed on the finite set $I$.  The \emph{nerve} of $\mathcal A$, denoted by \(\Nerve(\mathcal A)\) is the simplicial complex
 \(\Nerve(\mathcal A)
 =\left\{\sigma\subseteq I:
          \bigcap_{i\in\sigma}A_i\ne\emptyset\right\}.\)
By convention, the empty set is included as a face in the nerve to satisfy the definition of a simplicial complex. 
\end{definition}

A non-empty set \(\sigma \subseteq I\) is considered a face of the nerve if and only if the subfamily of \(\mathcal{A}\) indexed by \(\sigma\) has a non-empty intersection. Note that the nerve of a family records intersections of arbitrary subfamilies, not merely
pairwise intersections.  

\begin{definition}
A finite simplicial complex $\kk$ is \emph{$d$-representable} if there is a
finite family $\mathcal A = \{A_i\}_{i \in I}$ of convex sets in $\mathbb R^d$ such that
\(
 \kk\cong\Nerve(\mathcal A),
\)
i.e., the faces of $\kk$ can be identified with the index sets of subfamilies of $\mathcal A = \{A_i\}_{i \in I}$
having a non-empty total intersection.
\end{definition}

\begin{definition}
    Let $\kk$ and $\mathsf{L}$ be two simplicial complexes with disjoint vertex sets. Then their \emph{simplicial join}, denoted by $\kk * \mathsf{L}$, is defined as follows: \(\kk*\mathsf{L}=\{\sigma\cup\tau:\sigma\in \kk,\ \tau\in \mathsf{L}\}.\) This means, every face of $\kk * \mathsf{L}$ is obtained by independently choosing a face of $\kk$ and
a face of $\mathsf{L}$. 
\end{definition}

From the above definition, it is easy to see that $\dim(\kk*\mathsf{L})=\dim \kk+\dim \mathsf{L}+1$, whenever both complexes have a vertex.

To design tools to establish lower bounds on the boxicity of power graphs, we examine a specific simplicial complex $D_3$, the zero-dimensional complex on three vertices:
\( D_3=\{\emptyset,\{1\},\{2\},\{3\}\}.\) For $r\ge 1$, we then take the simplicial join $\dd$ of $r$ disjoint copies of $D_3$. Note that a face of $\dd$ has at most one of the three vertices from each copy of $D_3$. Moreover, a maximal face contains one vertex from each of the $r$ copies of $D_3$ and hence each maximal face has $r$ vertices. So, $\dim \dd = r-1$. 

We use the following theorem by Tancer \cite{Tancer_2011} as a black box in our paper.

\begin{theorem}[\cite{Tancer_2011}]\label{thm: original tancer}
   If $\kk$ is an $s$-dimensional simplicial complex with a non-vanishing Van Kampen obstruction, then the barycentric subdivision $\sd \kk$ of $\kk$ is not $2s$-representable.
\end{theorem}

The \emph{Van Kampen obstruction} is a certain obstruction for the embeddability of $d$-dimensional simplicial complexes into $\mathbb R^{2d}$. Defining this obstruction precisely would require extensive preliminaries, and since no definition or calculation is necessary later, we avoid mentioning the details here and direct interested readers to \cite[Section~2]{bestvina2002van} and \cite[Appendix~D]{matousek2011hardness} for an introductory explanation. From the discussion in \cite[Section~2.2 and Lemma~9]{bestvina2002van}, it is clear that the $\dd$ has a non-vanishing Van Kampen obstruction in target dimension $2r-2$. Combining this fact with \Cref{thm: original tancer}, one can derive the following corollary easily.



\begin{corollary}
\label{thm:tancer}
For every $r\ge2$, the barycentric subdivision
\(\sd(D_3^{*r})\) is not $(2r-2)$-representable.
\end{corollary}

The above corollary serves as the primary tool for establishing the lower bounds discussed in this paper.

\subsection{
The Comparability Graph of the Face Poset of $\dd$}
\label{subsec:boxes-to-nerves}

In this subsection, we establish a lower bound for the boxicity of the comparability graph of the face poset of $\dd$. The proof utilises \Cref{thm:tancer}. Observe that \Cref{thm:tancer} concerns nerves of convex sets and boxicity is
defined using axis-parallel boxes, which are examples of convex sets. We now discuss why an axis-parallel box representation produces exactly the nerve needed above.

Before going into further details, we state the \emph{$2$-Helly property for axis-parallel boxes}: Let $\mathcal S$ be a nonempty finite family of axis-parallel boxes in $\mathbb R^d$.  Then \(\bigcap_{S\in\mathcal S}S\ne\emptyset \Longleftrightarrow
 S\cap S'\ne\emptyset \text{ for every }S,S'\in\mathcal S.\)
Note that the preceding statement is stronger than Helly's theorem for
general convex sets in $\mathbb R^d$.  It holds because an axis-parallel box is a Cartesian
product of intervals, and Helly's theorem applies for intervals in $\mathbb R$.

We can now derive the main result of this subsection.

\begin{lemma}
\label{lem:ternary-nerve}
Suppose $\G_r=\Comp(\operatorname{P}(\dd))$. For every $r\ge1$,
$
\boxi(\G_r)\ge2r-1.
$
\end{lemma}

\begin{proof}
For $r=1$, it is easy to see that $\G_1$ is the empty graph on three vertices. Hence, $\boxi(\G_1)=1$.  

Now assume $r\ge2$.  For a contradiction, suppose that
\(\boxi(\G_r)\le2r-2.\) So we consider a $(2r-2)$-box representation
\(\mathcal{B}=\{B_\sigma: \ \sigma \in \Pt(\dd)\}\) of $\G_r$, where each $B_{\sigma}$ is an axis-parallel box in $\mathbb R ^{2r-2}$. Note that a vertex of $\G_r$ is an element of the poset $\Pt(\dd)$ and so a non-empty face of $\dd$. 

We now take the nerve of the family \(\mathcal B =\{B_\sigma: \ \sigma \in \Pt(\dd)\}\). We first prove the following claim:

\begin{claim}\label{claim:nerve and barycentric subdivision}
    $\Nerve (\mathcal{B})=\Delta(\Pt(\dd))=\sd \dd$.
\end{claim}

\begin{proof}
    Let us take a face $F=\{\sigma_1,\sigma_2,\dots,\sigma_t\}$ from $\Nerve(\mathcal{B})$. We show that this face $F$ corresponds to a face in $\Delta(\Pt(\dd))$, i.e., to a finite chain in $\Pt(\dd)$. Note that $F=\{\sigma_1,\sigma_2,\dots,\sigma_t\} \subseteq \Pt(\dd)$ and $\bigcap_{i=1}^t B_{\sigma_i} \neq \emptyset$. This means that $\{\sigma_i\}_{i=1}^t$ forms a clique in $\G_r$ and $\{\sigma_i\}_{i=1}^t$ forms a chain (in some order) in $\Pt(\dd)$. This, in turn, implies that $\{\sigma_i\}_{i=1}^t$ is a face in $\Delta(\Pt(\dd))=\sd \dd$.

    For the opposite direction, we consider a face $F'=\{\tau_i\}_{i=1}^s$ in $\Delta(\Pt(\dd))$. This corresponds to a finite chain in $(\Pt(\dd))$. Our goal is to show that there is a corresponding face $F$ in $\Nerve(\mathcal{B})$. By construction, $\{\tau_i\}_{i=1}^s$ induces a clique in $\G_r$. So, $B_{\tau_i}\cap B_{\tau_j}\neq \emptyset$, for each $1\leq i < j \le s$. By Helly's property of axis-parallel boxes, we have $\bigcap_{i=1}^sB_{\tau_i}\neq \emptyset$. Hence, the set $\{\tau_i\}_{i=1}^s$ is a face in $\Nerve(\{B_{\sigma} \ : \ \sigma \in \Pt(\dd)\})$.
\end{proof}

Since an axis-parallel box is a convex set, the above claim implies that $\mathcal{B}$ is a convex set representation of $\sd \dd$ in
$\mathbb R^{2r-2}$; equivalently, $\sd \dd$ is
$(2r-2)$-representable.  This contradicts \Cref{thm:tancer} and therefore, our assumption is wrong and hence \( \boxi(\G_r)>2r-2\).
\end{proof}

Note that the equality between the nerve and the barycentric subdivision mentioned in \Cref{claim:nerve and barycentric subdivision} does not hold in general. This is because pairwise intersections need not always guarantee a common point. However, in this context, equality holds because we are dealing with a box representation of a comparability graph, and the Helly property for axis-parallel boxes guarantees the required result.

%% file: comparability_graph_box_tool.tex
In this section, we develop some tools that will eventually help us bound the boxicity of a particular product of posets by breaking it down into more basic units, such as chains and rooted tree posets.

\subsection{The comparability graph of a rooted tree poset}

\begin{lemma}\label{lem:dfs}
Let $T$ be a finite rooted tree poset.  Then $\Comp(T)$ is an
interval graph.
\end{lemma}

\begin{proof}
We first traverse the cover graph of $T$ in depth-first order and list its leaves from left to right during the traversal.  For every vertex $v$, let $l(v)$ and $r(v)$ be the positions in the depth-first order of the first and last leaves below $v$, possibly including $v$ if it is itself a leaf. Assign the interval $J(v)=[l(v),r(v)]$. All leaves below a fixed node occur consecutively in a depth-first order. Note that if $v$ is a leaf, then $l(v)=r(v)$ and $J(v)$ is a point interval. 

Let $u$ be an ancestor of $v$. If $v$ is a leaf, then clearly $J(v)\subseteq J(u)$. If $v$ is a non-leaf vertex, then every leaf below $v$ is also below $u$, so
$J(v)\subseteq J(u)$.  If $u$ and $v$ are incomparable, their rooted
subtrees are disjoint and occupy disjoint blocks of leaves, so
$J(u)\cap J(v)=\varnothing$.  Thus, intervals meet exactly for
ancestor-comparable nodes. Therefore, the intersection graph of these intervals is isomorphic to the interval graph $\Comp(T)$.
\end{proof}

\subsection{Adding a chain coordinate}


Recall that $C_e$ denotes a chain $\{0<1<2<\dots<e\}$ of length/height $e$.

\begin{lemma}\label{lem:add-chain}
For every finite poset $P$ and every $e\ge0$, $\boxi(\Comp(P\times C_e)) \le \boxi(\Comp(P))+e$.
\end{lemma}

\begin{proof}
Let $d=\boxi(\Comp(P))$ and let $J_1, J_2, \ldots , J_d$ be a collection of interval graphs that form a $d$-box representation of
$\Comp(P)$. 

Consider an arbitrary linear extension of $P$ and for an element $x\in P$, let $\lambda(x)$ be the position of $x$ in the linear extension. Let $r\in P$ be the element appearing last in this linear order.

We define interval graphs $I_1, I_2, \ldots , I_{d+e}$ such that $\forall (x,j)\in P\times C_e$, we have $$\forall i\in [d], I_i(x,j)=J_i(x)$$ and $$\forall i\in [d+e]\setminus [d], I_i(x,j)=\begin{cases}
  [\lambda(x),\lambda(r)],&j<i-d,\\
  [0,\lambda(x)],&j\ge i-d.
 \end{cases}$$

Note that when two elements $x$ and $y$ of $P$ are incomparable, there exists $k\in [d]$, such that $J_k(x)\cap J_k(y)=\emptyset$. Consequently, for any $j_1,j_2\in C_e$, $I_k(x,j_1)\cap I_k(y,j_2)=\emptyset$.

Therefore, we only need to check the case when two elements $x$ and $y$ of $P$ are comparable.

Suppose $(x,j_1)\le(y,j_2)$ in the product, so $x\le y$ and $j_1\le j_2$.
It is clear that for $i\in [d], I_i(x,j_1)\cap I_i(y,j_2)\neq \emptyset$.  For $i\in [d+e]\setminus [d]$, there are three possibilities:
\begin{itemize}
\item $j_1,j_2<i-d$: we have $\lambda(r)\in I_i(x,j_1)\cap I_i(y,j_2)$;
\item $i-d\le j_1,j_2$: we have $0\in I_i(x,j_1)\cap I_i(y,j_2)$;
\item $j_1<i-d\le j_2$: the intervals are $[\lambda(x),\lambda(r)]$ and
$[0,\lambda(y)]$, which meet because $\lambda(x)\le\lambda(y)$.
\end{itemize}
Hence, every such pair remains adjacent.

Now suppose $x<y$ but $j_1>j_2$.  Choose $k$ with $j_2<k\le j_1$.  Then $I_{d+k}(x,j_1)=[0,\lambda(x)],  I_{d+k}(y,j_2)=[\lambda(y),\lambda(r)]$, and these intervals are disjoint because $\lambda(x)<\lambda(y)$.  The case $y<x$ and $j_2>j_1$ is symmetric.  These are precisely the remaining pairs in the product.
\end{proof}

\subsection{Replacing a chain by a rooted tree poset}

\begin{lemma}\label{lem:tree-refinement}
Let $P$ be a finite poset and let $T$ be a rooted tree poset of height $e$.  Then
$ \boxi(\Comp(P\times T)) \le \boxi(\Comp(P\times C_e))+1$.
\end{lemma}

\begin{proof}

Take a box representation of $\Comp(P\times C_e)$, where the vertex $(x,j)$ is given the box $\tilde{B}_{x,j}$. Assign the vertex $(x,v)\in P\times T$ the box $B_{x,v}\colon=\tilde{B}_{x,\depth(v)}$. By the proof of Lemma~\ref{lem:dfs}, there is an interval representation $J$ of $T$ such that for a vertex $v\in T$, the interval $J(v)$ is determined by the depth-first order of the leaves of $v$. 

We will show that the intersection graph of the boxes $\{B_{x,v}\times J(v)\}_{(x,v)\in P\times T}$ is isomorphic to $\Comp(P\times T)$. 

Suppose $(x,u), (y,v)\in P\times T$ are comparable. Then $(x,u)\leq (y,v)$ and so $(x,\depth(u))\leq (y,\depth(v))$. Hence, $B_{x,u}\times J(u)\cap B_{y,v}\times J(v)\neq \emptyset$

Conversely, suppose $(x,u), (y,v)\in P\times T$ are incomparable.  If
$(x,\depth(u))$ and $(y,\depth(v))$ are incomparable in
$P\times C_e$, then $\tilde{B}_{x,\depth(u)} \cap \tilde{B}_{y,\depth(v)}= \emptyset$ and consequently, $B_{x,u} \cap B_{y,v}= \emptyset$.  On the other hand, if $(x,\depth(u))$ and $(y,\depth(v))$ are comparable in $P\times C_e$, then the incomparability of $(x,u)$ and $(y,v)$ forces $u$ and $v$ to lie on different branches of $T$. In that case $J(u)\cap J(v)=\emptyset$. In either case, $B_{x,u}\times J(u)\cap B_{y,v}\times J(v)= \emptyset$
\end{proof}

%% file: some_particular_poset_structure.tex
\subsection{The Comparability Graph of Product of Star Posets}

Recall that the star poset $\mathrm{S}$ is defined as follows: $\mathrm{S}$ is the four-element star poset consisting of a root element $\rho$ and
three leaves $a_1,a_2,a_3$, where the relations in the poset are given by \( \rho<a_j, \ j=1,2,3\), with the leaves pairwise incomparable.  This poset is the face poset
of $D_3$ when the empty face is retained: the root $\rho$ of $\mathrm{S}$ refers to the
empty face of $D_3$, and each leaf $a_j$ refers to the singleton face $\{j\}$. Note that by this definition of $\mathrm{S}$, the product poset $\mathrm{S}^r$ explicitly corresponds to the full face poset of $\dd$, i.e., the face poset of $\dd$ including the empty face as well. Hence, the following observation is straightforward. The proof is included in \Cref{app: proof of H_r induced subgraph}

\begin{observation}\label{obs: H_r is an induced subgraph}
    $\G_r=\Comp(\Pt(\dd))$ is an induced subgraph of $Comp(\mathrm{S}^r)$. This in turn implies that $\boxi(\Comp(\mathrm{S}^r)) \geq 2r-1$.
\end{observation}


\subsection{The Comparability Graph of a Mixed Product of Chains and Star Posets}
\label{subsec:boolean-product}

Let $C_1^r$ denote the product of $r$ copies of the poset chain \( C_1=\{0<1\}\). Hence, an element of $C_1^r$ is a tuple
\(
 \mathbf{x}=(x_1,\ldots,x_r),\ x_i\in\{0,1\},
\)
and the product order is coordinate-wise:
\(
 \mathbf{x}\le \mathbf{y}
 \text{ if and only if }
 x_i\le y_i\text{ for every }i.
\) The map \( x\mapsto\{i\in[r]:x_i=1\}\) is an order isomorphism from $C_1^r$ to the \emph{Boolean lattice} of all subsets of $[r]$, ordered by inclusion. 
This comparability graph $\Comp(C_1^r)$ was referred to in \cite{chandran2026boxicity} as the \emph{transitive closure of the hypercube of dimension $r$}, denoted by the notation $TC(H_r)$. They used the function $f(r)$ to represent the boxicity $\boxi(TC(H_r))$ and we retain this notation, i.e., we define $f(r)=\boxi(TC(H_r))=\boxi(\Comp(C_1^r))$.

The following observation is stated in \cite[Corollary~3]{chandran2026boxicity} in terms of $TC(H_r)$. But we note here for completeness, as it is essential for our discussions in this section and throughout the remainder of the paper. 
This observation is easy to see from combining the fact $\boxi(\Comp(C_1))=0$ and \Cref{lem:add-chain}. 

\begin{observation}\label{obs: boolean box upper bound}
    $\boxi(\Comp(C_1^r))\leq r-1$, $r\ge 1$.
\end{observation}

\begin{proof}
    The comparability graph $\Comp(C_1)$ is complete and, under our convention, \(f(1)=0.\) For $r\ge2$, write \( C_1^r=C_1^{r-1}\times C_1\).
Lemma~\ref{lem:add-chain} gives
\(f(r)\le f(r-1)+1\), since the height of the chain is $1$. Iterating this inequality from $f(1)=0$ yields $f(r)\le r-1$.
\end{proof}

\noindent \textit{A poset construction:} We consider the product of posets \(R_{r,j} = \mathrm{S}^j \times C_1^{r-j}\) for \(j = 0, 1, \ldots, r\). In particular, this is a product of multiple copies of two types of posets: $j$ copies of a star poset $\mathrm{S}$ and $ r-j$ copies of a chain. It is important to note that \(R_{r,0} = C_1^r\) and \(R_{r,r} = \mathrm{S}^r\). 

In the next lemma (\Cref{lem: box of general R_j}), we will demonstrate the boxicity of the comparability graph of the poset \(R_{r,j}\) for \(0 \leq j \leq r\). 

\begin{lemma}\label{lem: box of general R_j}
 For $r\geq 1$, $\boxi(\Comp(R_{r,j}))=r+j-1$, for $0\le j \le r$.
\end{lemma}
\begin{proof}
   The strategy is to go from $R_{r,j-1}$ to $R_{r,j}$ by replacing one factor $C_1$ with $\mathrm{S}$ and applying \Cref{lem:tree-refinement}. Both $C_1$ and $\mathrm{S}$ have height 1, and hence using \Cref{lem:tree-refinement} we have the following inequality: \(\boxi(\Comp(\mathrm{S}^{j-1} \times C_1^{r-j}\times \mathrm{S})) \le \boxi(\Comp(\mathrm{S}^{j-1} \times C_1^{r-j}\times C_1))+1\). Since permuting the positions of the factors does not change the final product up to isomorphism, the above inequality can be rewritten as follows: 
   \begin{equation}\label{eq: r_j}
       \boxi(\Comp (R_{r,j})) \le \boxi(\Comp(R_{r,j-1}))+1. 
   \end{equation}

By repeatedly applying \Cref{lem:tree-refinement} to the right-hand side of the inequality in \Cref{eq: r_j} until we reach \(j=0\) (which requires applying \Cref{lem:tree-refinement} a total of \(j\) times), we obtain the following result: \(\boxi(\Comp(R_{r,j})) \le \boxi(\Comp(C_1^{r})) + j\). Next, by using \Cref{obs: boolean box upper bound}, we arrive at the desired upper bound: \(\boxi(\Comp(R_{r,j})) \le r - 1 + j\).

For the lower bound, we apply \Cref{eq: r_j} successively for $\ell=j+1,j+2,\dots,r$. Thus, we have $\boxi(\Comp(R_{r,r}))\le \boxi(\Comp(R_{r,j}))+r-j$. Since $R_{r,r}=\mathrm{S}^r$, by \Cref{obs: H_r is an induced subgraph} the last mentioned inequality yields the following: $ 2r-1 \le \boxi(\Comp(R_{r,j}))+r-j$. Thus, we have $\boxi(\Comp(R_{r,j})) \ge r+j-1$.
\end{proof}

As a direct consequence of the previous lemma, we will now determine the exact boxicity of the comparability graph of the Boolean lattice \(C_1^r\). As discussed earlier, this result also provides the precise value of the boxicity of the power graph of the cyclic group, which was an open question in \cite{chandran2026boxicity}.

\begin{corollary}
\label{lem:boolean}
For every $r\ge1$,
$
 f(r) = \boxi(\Comp(C_1^r)) = r-1.
$
\end{corollary}

\begin{proof}

Putting $j=0$ in \Cref{lem: box of general R_j}, we have the lemma.
\end{proof}

 To ensure uniformity in exposition, the proof of \Cref{lem:boolean} has been presented in its current form as a consequence of~\Cref{lem:ternary-nerve}. However, there is a more direct and combinatorial proof of this lemma, which may be of independent interest. We present this alternate proof in the Appendix~\ref{appendix: proof lemma 4.18}.

 \subsection{The Comparability Graph of Product of Some Special Types of Rooted Tree Posets}

In this subsection, we establish a lower bound for the boxicity of the comparability graph of a product of posets, in particular a product of rooted tree posets with special properties. The proof of this lemma uses the strategy for establishing the lower bound in \cite{chandran2026boxicity}.

 \begin{lemma}\label{lem:scheduled-forks}
Let \(d\ge1\), and let \(T_1,\ldots,T_d\) be finite rooted tree posets of heights \(e_1,\ldots,e_d\), respectively, where \(e_i\ge1\) for every \(i\in[d]\).  Suppose
that for every $i$ in a specified set $R$, the rooted tree poset $T_i$ 
satisfies $h_3(T_i)=e_i-1$.  Then 
\begin{equation}\label{eq:scheduled-forks}
 \boxi\!\left(\Comp\!\left(\prod_{i=1}^dT_i\right)\right)
 \ge \sum_{i=1}^de_i-\max_i e_i+|R|.
\end{equation}
\end{lemma}

\begin{proof}
Put $M=\max_i e_i$.  We first choose a small induced subposet $P_i$ in every factor rooted tree poset $T_i$.  
For \(i\notin R\), choose in \(T_i\) a root-to-leaf chain \(P_i\) of height
\(e_i\), so \(P_i\cong C_{e_i}\). For \(i\in R\), choose a node \(c_i\) of
depth \(e_i-1\) having three pairwise incomparable strict descendants; these descendants have depth \(e_i\), since
\(T_i\) has height \(e_i\). In this case, i.e., for $i \in R$, let \(P_i\)
consist of the root-to-\(c_i\) chain together with three such descendants. Thus \(P_i\) is an induced subposet of \(T_i\), and consequently
\(\prod_{i=1}^d P_i\) is an induced subposet of \(\prod_{i=1}^d T_i\).  It is therefore enough to prove the required lower bound for
$\Comp(\prod_iP_i)$.

We consider the dual order
$
 Q=\left(\prod_{i=1}^dP_i\right)^{\mathrm{op}}
   =\prod_{i=1}^dP_i^{\mathrm{op}}.
$ Since the dual order $P^{\mathrm{op}}$ of a poset $P$ is defined to be the same set, but with the inverse order, i.e. $x \le y$ holds in $P^{\mathrm{op}}$ if and only if $y \le x$ holds in $P$, taking the duality does not change the comparability graph. We list the different $e_i+1$ levels of a subposet $P_i^{\mathrm{op}}$ from bottom to top as $L_{i,0}, L_{i,1}, \ldots, L_{i,e_i}$. For $i\notin R$, these levels of $P_i^{\mathrm{op}}$ are singleton sets. 
For $i\in R$, the bottom most level $L_{i,0}$ has three elements, every $L_{i,j}$ with $1\le j\le e_i$ is a singleton, and every element of $L_{i,0}$ lies below the element of $L_{i,1}$.  In either
case, every element at a lower-indexed level lies below every element at a
higher-indexed level.

For $1\le k\le M$, we set
$A_k=\{i:e_i\ge k\}$ and $r_k=|A_k|.$ Since every $e_i\ge1$, we have $A_1=\{1,\ldots,d\}$ and $r_1=d$. Also note that, $A_M \subseteq A_{M-1}\subseteq \dots \subseteq A_2 \subseteq A_1$. Let $V_k\subseteq Q$ consist of the tuples $\mathbf{x}=(x_1,\ldots,x_d)$ satisfying
\[
 \begin{cases}
 x_i\in L_{i,k-1}\cup L_{i,k},&i\in A_k,\\
 x_i\in L_{i,e_i},&i\notin A_k,
 \end{cases}
\]
Let \(m_k\) be the unique maximum of \(V_k\), obtained by taking the element
of \(L_{i,k}\) when \(i\in A_k\) and that of \(L_{i,e_i}\) otherwise, and put
\(V_k^*=V_k\setminus\{m_k\}.\)

We use the elementary fact that deleting a universal vertex does not change
boxicity. After permuting the factors,
\((V_1)^{\mathrm{op}}\cong \mathrm{S}^{|R|}\times C_1^{\,d-|R|}=R_{d,|R|}\) (See the construction of $R_{r,j}$ in \Cref{subsec:boolean-product}.). Under this isomorphism, \(m_1\) corresponds to the unique minimum of \(R_{d,|R|}\), which is universal in its comparability graph.  Therefore \Cref{lem: box of general R_j} gives
\begin{equation}\label{eq:first-dual-fork-layer}
 \boxi(\Comp(V_1^*))=d+|R|-1.
\end{equation}

For \(k\ge2\), both levels \(L_{i,k-1}\) and \(L_{i,k}\) are singletons for
every \(i\in A_k\). Hence
\( (V_k)^{\mathrm{op}}\cong C_1^{\,r_k}.\) Since \(V_k^*\) is obtained by deleting the unique maximum of \(V_k\), \Cref{lem:boolean} gives
\begin{equation}\label{eq:later-dual-boolean-layers}
 \boxi(\Comp(V_k^*))=r_k-1
 \qquad(2\le k\le M).
\end{equation}

We now verify that these layers form an induced join.
\begin{claim}
   The subgraph induced by $\bigcup_{k=1}^M V^*_k$ is $\Comp(V^*_1)\vee\Comp(V^*_2)\vee\cdots\vee\Comp(V^*_M).$ 
\end{claim}
\begin{proof}
    The vertex sets $V^*_1,V^*_2,\dots,V^*_M$ are pairwise disjoint. If \(l\ge k+2\), choose \(i\in A_l\); the \(i\)-th coordinates of \(V_k\) and \(V_l\) lie in disjoint sets of
levels. If \(l=k+1\), a tuple belonging to both \(V_k\) and \(V_{k+1}\) must lie in \(L_{i,k}\) for every \(i\in A_k\) and in \(L_{i,e_i}\) for every \(i\notin A_k\). It is therefore \(m_k\), which does not belong to \(V^*_k\).

Now let \(k< l\), \(x\in V^*_k\), and \(y\in V_l\). For each \(i\), if
\(e_i< l\), then \(y_i\) is the maximum element of \(P_i^{\mathrm{op}}\),
so \(x_i\le y_i\). If \(e_i\ge l\), then
\(x_i\in L_{i,k-1}\cup L_{i,k},\ y_i\in L_{i,\ell-1}\cup L_{i,\ell},\)
and \(k\le l-1\), so again \(x_i\le y_i\). Hence \(\mathbf{x}\le \mathbf{y}\) in \(Q\).
Thus the subgraph of $\Comp(Q)$ induced by the set \(\bigcup_{k=1}^M V^*_k\) is the join \(\bigvee_{k=1}^M\operatorname{Comp}(V^*_k).\)
\end{proof}

By Lemma~\ref{lem:join}, \eqref{eq:first-dual-fork-layer}, and
\eqref{eq:later-dual-boolean-layers},
\begin{align*}
 \boxi(\Comp(Q))
 &\ge d+|R|-1+\sum_{k=2}^M(r_k-1)\\
 &=\sum_{k=1}^M r_k-M+|R|\\
 &=\sum_{i=1}^d e_i-M+|R|.
\end{align*}

The last equality holds because \(i\in A_k\) for exactly \(e_i\) values of
\(k\).  Finally, $\Comp(Q)=\Comp(\prod_iP_i)$ is an induced subgraph of
$\Comp(\prod_iT_i)$, which proves the lemma.
\end{proof}

\subsection{The Comparability Graph of Mixed Product of a Chain and a Special type of Rooted Tree Poset}

For later use, observe that if $T$ is any rooted tree poset of height $b$, then
Lemma~\ref{lem:tree-refinement} and \Cref{lem:add-chain} give
\begin{equation}\label{eq:basic-tree-chain-bound}
 \boxi(\Comp(T\times C_n))\le \min\{b,n\}+1.
\end{equation}

\begin{lemma}
\label{lem:low-branching-absorption}
Let $T$ be a finite rooted tree poset of height $a\ge1$, let $m\ge1$, and put
\[
 q=\min\{a,m\}.
\]
If $h_3(T)\leq q-2$ and $h_2(T)\leq q-1$, then
\begin{equation}\label{eq:low-branching-absorption}
\boxi(\Comp(T\times C_m))=q.
\end{equation}
\end{lemma}

\begin{proof}
Choose a leaf of maximum depth $a$, and let $C_a$ be the induced chain of its ancestors, including the leaf. Its product with $C_m$
is an induced $C_a\times C_m$, whose comparability graph has boxicity at least $q$ by
\Cref{thm:product-chains}.  This proves the lower bound.

We now construct a $q$-box representation by induction on $q$.

\medskip
\noindent
\emph{Base case 1: $q=1$ and $a=1\le m$.}
The rooted tree poset $T$ consists of its root $\rho$ and at most two leaves because $h_3(T)\le q-2$ forbids three incomparable descendants of the root. If $T$ consists of only one leaf, $T=C_1$ and so, $\boxi(\Comp(T\times C_m))=1=q$. Therefore, assume that $T$ consists of exactly 2 leaves, say $x$ and $y$. 

For $0\le i,j\le m$, assign
\begin{align*}
 I(\rho,i)&=[i,2m+2-i],\\
 I(x,j)&=[-1,j],\\
 I(y,j)&=[2m+2-j,2m+3].
\end{align*}
We show that this is an interval representation of $\Comp(T\times C_m)$. 

Suppose two of its vertices $(u_1,j_1)$ and $(u_2,j_2)$ are adjacent; then they form a chain of length 1. Without loss of generality, assume $u_1\leq u_2$ and $j_1\leq j_2$. If $u_1=x$ then $u_2$ must be $x$. It is clear that $-1\in I(x,j_1)\cap I(x,j_2)$. Similarly, if $u_1=y$ then $u_2$ must be $y$ and so, $2m+3\in I(y,j_1)\cap I(y, j_2)$. Therefore, assume $u_1=\rho$. If $u_2=\rho$, then  $m+1\in I(\rho,j_1)\cap I(\rho, j_2)$. If $u_2=x$, then $ I(\rho,j_1)\cap I(x, j_2)\neq \emptyset $ as $j_1\leq j_2$. Similarly, if $u_2=y$, then $ I(\rho,j_1)\cap I(y, j_2)\neq \emptyset $ as $2m+2-j_2\leq 2m+2- j_1$ (since $j_1\leq j_2$). 

Now suppose $(u_1,j_1)$ and $(u_2,j_2)$ are not adjacent. Then it must be that they are incomparable. If \(\{u_1,u_2\}=\{x,y\}\), the corresponding intervals are
disjoint, since every \(x\)-interval ends at most at \(m\),
whereas every \(y\)-interval starts at least at \(m+2\).
Otherwise, without loss of generality, assume that
\(u_1<u_2\) and \(j_1>j_2\). This forces \(u_1=\rho\). If $u_2=x$, $I(\rho,j_1)\cap I(x,j_2)=\emptyset$ as $j_1>j_2$. If $u_2=y$, $I(\rho,j_1)\cap I(y,j_2)=\emptyset$ as $2m+2-j_2>2m+2-j_1$ (since $j_1>j_2$).

\medskip
\noindent
\emph{Base case 2: $q=1$ and $m=1<a$.}
$h_2(T)\le q-1$ forbids two-way branching at every positive depth,
while property $h_3(T)\le q-2$ permits at most two branches from the root.
Consequently, $T$ is one root-to-leaf chain or the union of two such chains
meeting only at the root.

Write $v_d$ for the vertex at depth $d$ on the first branch and $w_d$ for
the vertex at depth $d$ on the second branch.  Put $K=a+2$ and define
\begin{alignat*}{2}
 I(\rho,0)&=[-K,K],&\qquad
 I(\rho,1)&=[-1,1]\\
 I(v_d,0)&=[-K,-d-1],&
 I(v_d,1)&=[-d-1,-1],\\
 I(w_d,0)&=[d+1,K],&
 I(w_d,1)&=[1,d+1].
\end{alignat*}
We will show that this is an interval representation of $\Comp(T\times C_m)$. 

If two vertices of $\Comp(T\times C_m)$ are adjacent then their first coordinate should be from the same branch of the $T$ (including the root). For convenience, we will take $v_0=w_0=\rho$. Let the two vertices adjacent in $\Comp(T\times C_m)$ be on the first branch, say $(v_{d_1},j_1)$ and $(v_{d_2},j_2)$. Then these two vertices form a chain of length 1. Without loss of generality, assume that $(v_{d_1},j_1)\leq (v_{d_2},j_2)$. If $j_1=1$, then $j_2=1$. It follows that $-1\in I(v_{d_1},1)\cap I(v_{d_2},1)$. Therefore, assume $j_1=0$. If $j_2=0$, then $-K\in I(v_{d_1},0)\cap I(v_{d_2},0)$. Otherwise $j_2=1$. If \(d_1=0\), then \((v_{d_1},0)=(\rho,0)\) is universal and hence we are done. So $d_1\geq 1$. This forces, $d_2\geq 1$. It follows that $I(v_{d_1},0)\cap I(v_{d_2},1)\neq \emptyset$ since $-d_2-1\leq -d_1-1$ (since $d_1\leq d_2$). A similar argument works when the first coordinate of the vertices belong to the second branch of $T$ as the endpoints of the intervals are symmetric around the origin. 

If two vertices of $\Comp(T\times C_m)$ are not adjacent then they are not comparable. The first case is when their first coordinate belongs to the same branch of $T$. Without loss of generality, suppose the vertices are of the kind $(v_{d_1},j_1)$ and $(v_{d_2},j_2)$ and suppose $d_1<d_2$, $j_1>j_2$. This forces $j_1=1$ and $j_2=0$. Since, $(\rho,0)$ is universal, we can assume $v_{d_2}\neq \rho$, i.e., $d_2\geq 1$. But then, $I(v_{d_1},1)\cap I(v_{d_2},0)=\emptyset$ as $-d_2-1< -d_1-1$. A similar argument works when the first coordinate of the vertices belong to the second branch of $T$ as the endpoints of the intervals are symmetric around the origin. The second case is when the first coordinate of one vertex is from the first branch and that of the other is from the second branch (none of them have first coordinate equal to $\rho$). In this case, observe that $\bigcup_{d\neq 0}I(v_d,0)\cup I(v_d,1)=[-K,-1]$ and $\bigcup_{d\neq 0}I(w_d,0)\cup I(w_d,1)=[1,K]$. Therefore, the two intervals corresponding to the two vertices do not intersect.

\emph{Induction step 1: $a\le m$.}
Here $q=a\ge2$.  Let $\rho$ be the root, and let
$T_1,\ldots,T_s$ be the rooted subtree posets whose roots are the children of
$\rho$.  Removing the root row from $T\times C_m$ leaves the disjoint union
\(
 \bigsqcup_{i=1}^s\Comp(T_i\times C_m).
\)
If $T_i$ has height at most $a-2$,
$\boxi(\Comp(T_i\times C_m))\leq a-1$ by \eqref{eq:basic-tree-chain-bound}.

On the other hand, if $T_i$ has height $a-1$, $h_2(T_i)\le a-2$ and $h_3(T_i)\le a-3$. The induction hypothesis represents
$\Comp(T_i\times C_m)$ in $a-1$ dimensions.

Lemma~\ref{lem:disjoint-union} therefore represents the
entire displayed disjoint union in $a-1$ dimensions. Let $I_2, I_3,\ldots,I_a$ be the set of interval graphs representing $\Comp(\bigsqcup_{i=1}^s(T_i\times C_m))$.

For every integer $0\leq k\leq m$, make the vertex $(\rho,k)$ universal in the interval graphs $I_2, I_3\ldots,I_a$.  Define interval graph $I_1$ by the assignment given as follows.
\[
 I_1(v,j)=\begin{cases}
     [0,j]& \text{ if } v\ne\rho,\\
 [j,m]& \text{ if } v=\rho.
 \end{cases}
\]
All nonroot intervals contain $0$, so this interval graph changes no adjacency
among the nonroot vertices.  Moreover, for $v\ne\rho$, $1\le k\le m$, and $0\le j\le m$
\[
 I_1(\rho,k)\cap I_1(v,j)\ne\varnothing
 \quad\Longleftrightarrow\quad k\le j,
\]
which is precisely the condition $(\rho,k)\le(v,j)$.  We obtain an
$a=q$ dimensional representation.

\emph{Induction step 2: $m<a$.}
Here $q=m\ge2$. Consider the following induced subgraph
\( H=\{(v,j):v\ne\rho,\ 1\le j\le m\}.\)
After replacing $j$ by $j-1$, it is the disjoint union
\(
H\cong\bigsqcup_{i=1}^s\Comp(T_i\times C_{m-1}).
\)

If $T_i$ has smaller height, $\Comp(T_i\times C_{m-1})$ uses at most $m-1$ dimensions by \eqref{eq:basic-tree-chain-bound}. On the other hand, if $T_i$ has height at least $m-1$, then $h_2(T_i)\leq m-2$ and $h_3(T_i)\leq m-3$.  Therefore, the induction hypothesis gives an $(m-1)$-box
representation.  

By Lemma~\ref{lem:disjoint-union}, we have $m-1$ interval graphs $I_2,I_3\ldots,I_m$ whose intersection give 
$H$.

For $2\leq i\leq m$ and $v\neq \rho$, define $I_i(v,0)=I_i(v,1)$.  For every integer \(k\) with \(0\le k\le m\), make the vertex
\((\rho,k)\) universal in each of the interval graphs
\(I_2,I_3,\ldots,I_m\).  Let $d(v)=\depth_T(v)$ and put
$K=a+1$. Define the interval graph $I_1$ by the following assignment
\[
 I_1(v,j)=\begin{cases}
     [-K,-d(v)] &\text{ if } v\neq \rho \text{ and }j=0;\\
[-d(v),K] &\text{ if } v\neq \rho\text{ and }j=1;\\
 [-K,jK] &\text{ if } v\neq \rho\text{ and }j\ge2;\\
 [-K,mK]&\text{ if } v=\rho\text{ and }j=0;\\
 [jK,mK]& \text{ if } v= \rho\text{ and }j\ge1.
 \end{cases}
\]

It only remains to check the adjacency and non-adjacency of pairs involving newly
attached vertices.
\begin{itemize}
\item Two vertices $(v,0),(w,0)$ with $v\neq \rho$ and $w\neq \rho$ are adjacent in $I_1$ as $-K\in I_1(v,0)\cap I_1(w,0)$.  Therefore, $(v,0)$ and $(w,0)$ are adjacent in $I_1\cap I_2\cap \cdots I_m$ if and only if $ (v,1)$ and $(w,1)$ are adjacent in  $I_2\cap I_3\cap \cdots I_m$ which is true if and only if $v$ and $w$ are comparable in $T$. This is exactly the requirement for the adjacency of $(v,0)$ and $(w,0)$.
\item Two vertices $(v,0)$ and $(w,1)$ with $v\neq \rho$ and $w\neq \rho$ are adjacent in $I_1$ if and only if $-d(w)\leq -d(v)$, i.e., $d(v)\leq d(w)$. Moreover, since $I_i(v,0)=I_i(v,1)$ for every $2\le i\le m$,
the vertices $(v,0)$ and $(w,1)$ are adjacent in
$I_2\cap\cdots\cap I_m$ if and only if $v$ and $w$ are
comparable in $T$. Combining this with the condition
$d(v)\le d(w)$ imposed by $I_1$, we conclude that $(v,0)$
and $(w,1)$ are adjacent in $I_1\cap\cdots\cap I_m$
if and only if $v\le w$. This is precisely the condition
for $(v,0)$ and $(w,1)$ to be comparable in $T\times C_m$.
\item If $j\ge2$, $v\neq \rho$ and $w\neq \rho$ the adjacency of $(v,0)$ and $(w,j)$ is completely determined by $I_2, I_3, \dots I_m$.
\item The vertex $(\rho,0)$ is universal. For $v\ne\rho$, $0\le j\le m$, and $1\le k\le m$, we have,
\[
 I_1(\rho,k)\cap I_1(v,j)\ne\varnothing
 \quad\Longleftrightarrow\quad k\le j.
\]
Moreover, $I_1(\rho,k_1)\cap I_1(\rho,k_2)\neq \emptyset $ for $0\leq k_1,k_2\leq m$.
\end{itemize}
Thus, $I_1, I_2, \dots I_m$ gives an $m=q$ dimensional representation. This completes the induction.
\end{proof}

%% file: cyclic_overgroups_odd_p_group.tex
For this, we establish a property of the cyclic subgroup poset of a noncyclic $p$-group with odd $p$ (in \Cref{lem: terminal fork for odd prime}), which is used to establish our lower bound in \Cref{cor:odd-cyclic2}. First, we recall the standard normalizer condition for finite $p$-groups, stated in the next lemma;
see, for example, ~\citep[Sec~6.1, Th~1]{DummitFoote}. Let $gAg^{-1} = \{gag^{-1} : a \in A\}$. The normalizer of  $A$ in $G$ to be the set $N_G(A) = \{g \in G : gAg^{-1} = A\}$.

\begin{lemma}\label{lem: normalizer condition}\cite{DummitFoote}
Let $G$ be a finite $p$-group and let $H<G$. Then
\(
H<N_G(H).
\)
\end{lemma}



We will now discuss the main observation (see \Cref{lem: terminal fork for odd prime}) of this subsection. For the argument concerning the non-abelian case, we refer to the following statement from Suzuki's classification theorem \cite[Theorem~4.1]{suzukibook}. Here, $G=\langle T \ | \ R\rangle$ denotes a generator-relation presentation of a group $G$, where the set $T$ generates the group $G$ and $R$ contains the relations that the elements of $T$ satisfy.

\begin{lemma}\label{suzuki classification}\cite{suzukibook}
Let $p$ be odd and $G$ be a finite nonabelian
$p$-group of order $p^n$ possessing a cyclic maximal subgroup. Then
\(G\cong M(p^n),\) where, for $n\geq 3$,
\(M(p^n)=\left\langle a,b\ \middle|\
a^{p^{n-1}}=b^p=1,\ b^{-1}ab=a^{1+p^{n-2}}\right\rangle .\)
\end{lemma}

 We acknowledge that the following lemma is not a new result, and while it may have been noted by some group theorists, we have not found it stated verbatim. Therefore, we present the proof here.

\begin{lemma}\label{lem: terminal fork for odd prime}
Let $p$ be an odd prime, and let $G$ be a finite noncyclic $p$-group of
exponent $p^e$. Then there exists a cyclic subgroup $U\leq G$ of order
$p^{e-1}$ that is contained in at least three distinct cyclic subgroups
of order $p^e$.
\end{lemma}

\begin{proof}
The case $e=1$ is trivial. Then every nonidentity element of $G$ has
order $p$. Distinct subgroups of order $p$ intersect only in the
identity, and each such subgroup contains exactly $p-1$ nonidentity
elements. Hence, the number of subgroups of order $p$ is
\(
\frac{|G|-1}{p-1}.
\)
Since $G$ is noncyclic, $|G|\geq p^2$, and therefore
\(
\frac{|G|-1}{p-1}\geq\frac{p^2-1}{p-1}=p+1\geq4.
\)
Thus the trivial subgroup $U=\{1\}$, of order $p^{e-1}=1$, is
contained in at least three cyclic subgroups of order $p$. This proves
the lemma when $e=1$.

For the rest of our discussion, we assume that $e\geq2$. Since $G$ is a finite group, $\exp(G)=p^e$ implies
that some element $g\in G$ has order exactly $p^e$. Let
\(
A=\langle g\rangle.
\)
Then $|A|=p^e$. Since $G$ is noncyclic, $A<G$. By \Cref{lem: normalizer condition}
\(
A<N_G(A).
\)
By the definition of the normaliser, $A\trianglelefteq
N_G(A)$, so $N_G(A)/A$ is a nontrivial finite
$p$-group. Hence, by Cauchy's theorem, it contains a subgroup
$K'$ of order $p$.

Let
\(
\pi:N_G(A)\longrightarrow N_G(A)/A, \text{ defined by } \pi(x)=xA,
\)
be the natural quotient homomorphism. Put
\(
K=\pi^{-1}(K').
\)
Since the preimage of a subgroup under a homomorphism is a subgroup,
$K\leq N_G(A)\leq G$. Moreover, $\ker(\pi)=A\leq K$, and the restriction
$\pi|_K:K\to K'$ is surjective with kernel $A$. The First
Isomorphism Theorem, therefore, gives \(K/A\cong K'.\)
Consequently,
\(
[K:A]=p
\ \text{and}\
|K|=p^{e+1}.
\)
This implies that $A$ is a maximal subgroup of $K$. Also, $A$ is cyclic by construction.

Moreover $\exp(K)\mid p^e$, because $K\leq G$. On the other hand,
$K$ contains $g$, whose order is $p^e$. Hence
\(\exp(K)=p^e.\) So $K$ is noncyclic.

We now analyse the possible structure of $K$. If $K$ is abelian, we can write its standard decomposition as
\(
K\cong \Z_{p^{\lambda_1}}\times\cdots\times \Z_{p^{\lambda_r}},\ 
\lambda_1\geq\cdots\geq\lambda_r\geq1.
\)
By our discussion so far, it is easy to see that $\lambda_1=e$, $r=2$ and $\lambda_2=1$, so
\(
K\cong \Z_{p^e}\times \Z_p.
\) If $K$ is nonabelian, then $K$ is a nonabelian $p$-group with a cyclic
maximal subgroup $A$. Since $p$ is odd, Suzuki's classification theorem (mentioned in \Cref{suzuki classification}) gives
\(
K\cong M(p^{e+1}).
\)

We will complete the proof in the following two cases.

\medskip

\noindent \emph{Case 1:} $K$ is abelian, i.e., $K \cong \Z_{p^e}\times \Z_p$. We choose generators $a$ and $b$ such that \(K=\langle a\rangle\times\langle b\rangle,\ o(a)=p^e,\ o(b)=p.\) For each $i\in\{0,1,\ldots,p-1\}$, define \(x_i=ab^{-i}\). 
Because $a$ and $b$ commute and $b^p=1$,
\(x_i^p=a^p.\) 

\medskip

\noindent \emph{Case 2:} $K$ is non-abelian, i.e., $K \cong M(p^{e+1})$. We use the presentation of $M(p^n)$ with $n=e+1$ stated in \Cref{suzuki classification} and put $q=1+p^{e-1}$. Note that here $M(p^{e+1})\cong \Z_{p^e} \rtimes \Z_p$. Moreover, we choose generators $a,b$ of $K$ satisfying the relations \(a^{p^e}=b^p=1, \ b^{-1}ab=a^q.\) The last relation yields $b^{-j}ab^j=a^{q^j}$, for $j\geq 0$.

For $i\in\{0,1,\ldots,p-1\}$, we again define \(x_i=ab^{-i}.\)
The following results from repeatedly applying the group operation of $M(p^{e+1})$.
\[
\begin{aligned}
x_i^p
&=(ab^{-i})^p\\
&=a\,(b^{-i}ab^i)\,(b^{-2i}ab^{2i})
  \cdots(b^{-(p-1)i}ab^{(p-1)i})b^{-pi}\\
&=a^{1+\sum_{j=1}^{p-1}q^{ji}},
\end{aligned}
\]

Since $e\geq2$, 
\(
q^{ji}\equiv1+ji\,p^{e-1}\pmod{p^e}.
\)
It follows that
\[
\begin{aligned}
1+\sum_{j=1}^{p-1}q^{ji}
&\equiv \sum_{j=0}^{p-1}\bigl(1+ji\,p^{e-1}\bigr)\pmod{p^e}\\
&=p+i p^{e-1}\frac{p(p-1)}2\\
&=p+i p^e\frac{p-1}{2}\\
&\equiv p\pmod{p^e}.
\end{aligned}
\]
The last step is valid because $p$ is odd, and hence $(p-1)/2$ is an integer. Since
$o(a)=p^e$, our calculations so far yield that \(x_i^p=a^p.\)

Thus, in both Case 1 and 2, we have obtained \(
x_i^p=a^p \ (0\leq i\leq p-1).\) Let us define \(U=\langle a^p\rangle\) and $X_i=\ang{x_i}$.
Since $o(a)=p^e$, we have
\(|U|=p^{e-1}.\) Also, \(U=\langle x_i^p\rangle\leq \langle x_i\rangle=X_i, \text{ for } 0\leq i\leq p-1.\) If $o(x_i)=p^t$, then $o(x_i^p)=p^{t-1}$; hence $t=e$ (since $x_i^p$ has order $p^{e-1}$). Equivalently, each $X_i$ is a cyclic subgroup of
order $p^e$ which contains the cyclic subgroup $U$ of order $p^{e-1}$. It remains to verify that these $p$ cyclic subgroups $X_i$ are distinct. 

In the abelian case, it is trivial to see that $U$ is central in $K$, i.e., $U \le Z(K)$. In the nonabelian case, note that \(b^{-1}a^pb=(b^{-1}ab)^p=(a^q)^p=a^{p(1+p^{e-1})}=a^{p+p^e}=a^p\). This implies that the generator $a^p$ of $U$ commutes with both generators $a,b$ of $K$. Hence, $U \le Z(K)$ in the nonabelian case as well. Thus, the quotient $K/U$ is well-defined and \(|K/U|=\frac{p^{e+1}}{p^{e-1}}=p^2.\) 

The element $aU$ has order $p$. Also, $b\notin\langle a\rangle$ in both Case 1 and 2, so $bU\neq U$ and $bU$ has order $p$. These two
elements $aU$ and $bU$ generate $K/U$, and they commute. This is immediate in the
abelian case. In the nonabelian case, \(b^{-1}ab=a^{1+p^{e-1}}\) and \(a^{p^{e-1}}=(a^p)^{p^{e-2}}\in U.\) Hence, \(b^{-1}abU=a^{1+p^{e-1}}U=aU.\) Thus \(aU\) and \(bU\) commute in \(K/U\). Consequently, $K/U$ is an elementary abelian group of order $p^2$;
that is, \(K/U\cong \Z_p\times \Z_p.\)

The elementary abelian group $K/U \cong \Z_p\times \Z_p$ can be viewed as a two-dimensional vector space over $\F_p$, with the correspondence \((aU)^r(bU)^s\longleftrightarrow (r,s),\ r,s\in\mathbb F_p\) using $aU$ and $bU$ as a basis. Note that \(X_i/U=\langle x_iU\rangle
=\langle (aU)(bU)^{-i}\rangle=\langle(1,-i)\rangle\) in the two-dimensional vector space $K/U$ over $\F_p$. If $X_i/U=X_j/U$, then $\langle(1,-i)\rangle=\langle(1,-j)\rangle$ and hence
$(1,-i)=\lambda(1,-j)$ for some non-zero scalar $\lambda\in\F_p^{\times}$. Comparing
the first coordinates gives $\lambda=1$, and then $i=j$ in $\F_p$.
Thus $X_0/U,X_1/U,\ldots,X_{p-1}/U$ are pairwise distinct subspaces. This implies \(X_0,X_1,\ldots,X_{p-1}\) are pairwise distinct cyclic subgroups of order $p^e$, and all of them
contain the subgroup $U$ of order $p^{e-1}$. Since $p$ is odd,
$p\geq3$ and the required conclusion follows.
\end{proof}


\paragraph{Literature note.}
The maximal-order coset calculation in the proof of Theorem~1 of
Sambale-Wellmann~\citep{SambaleWellmann} gives an independent route to the
same conclusion.  The ternary fork formulation above is the specialisation
needed here and is proved explicitly to identify the common subgroup $U$.

%% file: lower_bound_proof_for_odd_p.tex

Now, using the above lemma and \Cref{lem:scheduled-forks}, we obtain the boxicity of the power graph of a nilpotent group with odd order or a nilpotent group that has cyclic Sylow 2-subgroups.

\begin{corollary}
\label{cor:odd-cyclic2}
Let $G$ be a finite nilpotent group. Then the following two hold: \\
(1) If the Sylow $2$-subgroup of $G$ is cyclic, then  $\boxi(\Pow(G))=\boxi(\Pow(\Z_{exp(G)}))+t(G)$.\\
(2) If $|G|$ is odd, then
$\boxi(\Pow(G))=\boxi(\Pow(\Z_{exp(G)}))+s(G)$.
\end{corollary}

\begin{proof}
Recall that \Cref{eq:central-reformulation} gives the following: $\boxi(\Pow(G))
 =\boxi\!\left(\Comp\!\left(\prod_i T_{i}\right)\right)$, where $T_{i}$ denotes the rooted tree poset arising from the Sylow $p_i$-subgroup $S_{p_i}$. In particular, $T_{i}$ is a chain exactly when $S_{p_i}$ is cyclic. Assuming $\exp(S_{p_i})=p_i^{e_i}$, height of the rooted tree poset $T_i$ is $e_i$. By \Cref{lem: terminal fork for odd prime}, every noncyclic Sylow subgroup $S_{p_i}$ of $G$ for odd $p_i$ satisfies $h_3(S_{p_i})=e_i-1$. There are $t(G)$ noncyclic odd order Sylow subgroups of $G$. Hence, applying \Cref{lem:scheduled-forks} we have the lower bound: $\boxi(\Pow(G))\ge \boxi(\Pow(\Z_{\exp(G)}))+t(G).
$

When the Sylow $2$-subgroup is absent or cyclic, the only noncyclic Sylow subgroups are the
$t(G)$ odd ones.  From the universal upper bound in \Cref{thm:universal-upper},  it follows that the two bounds coincide. Also, when $|G|$ is odd or Sylow $2$-subgroup is absent, then $t(G)=s(G)$.
\end{proof}